\documentclass[aps,prb,floatfix,reprint,superscriptaddress]{revtex4-2}
\usepackage{amsmath,amssymb}
\usepackage{graphicx}
\usepackage{epsfig}
\usepackage{psfrag}
\usepackage[usenames]{color}
\usepackage{hyperref}
\usepackage{soul,color}
\usepackage{tabu}
\usepackage{multirow}
\usepackage{bbold}
\usepackage{mathtools}
\usepackage{siunitx}
\usepackage{verbatim}
\usepackage{bbm}
\usepackage{enumitem}
\usepackage{braket}
\usepackage{appendix}
\usepackage{booktabs}
\usepackage{chngcntr}

\begin{document}
\title{Dynamics and lifetime of geometric excitations in moir\'e systems}

\author{Yuzhu Wang}
\affiliation{School of Physical and Mathematical Sciences, Nanyang Technological University, 639798 Singapore}

\author{Joe Huxford}
\affiliation{Department of Physics, University of Toronto, Toronto, Ontario M5S 1A7, Canada}

\author{Dung Xuan Nguyen}
\affiliation{Center for Theoretical Physics of Complex Systems, Institute for Basic Science (IBS), 34126 Daejeon, Korea}
\affiliation{Basic Science Program, Korea University of Science and Technology (UST), 34113 Daejeon, Korea}

\author{Guangyue Ji}
\affiliation{Department of Physics, Temple University, Philadelphia, Pennsylvania, 19122, USA}

\author{Yong Baek Kim}
\affiliation{Department of Physics, University of Toronto, Toronto, Ontario M5S 1A7, Canada}

\author{Bo Yang}
\affiliation{School of Physical and Mathematical Sciences, Nanyang Technological University, 639798 Singapore}
\date{\today}

\begin{abstract}
We show that spin-2 geometric excitations, known as graviton modes, generally exhibit vanishing lifetimes in lattice Chern bands, including in moir\'e systems. In contrast to the Landau levels, we first numerically demonstrate that the prominent graviton peaks in spectral functions diminish rapidly with increasing system sizes. We explore how the choice of interaction affects the strength of these peaks, with short-ranged interactions pushing the graviton mode far into the continuum of excitations, where it can be significantly scattered due to the increased density of states. We also analytically investigate the short lifetime of the graviton mode. In lattice systems, continuous rotational symmetry is broken, leading to highly anisotropic gapped excitations that mix different angular momentum or ``spins''. This is despite the surprising emergence of a ``guiding center" continuous rotational symmetry in the ground state, which is shared by the graviton mode. Consequently, the graviton mode in Chern bands can be strongly scattered by the anisotropic gapped excitations. However, the emergent rotational symmetry implies that gravitons can be robust in principle, and we propose experimental tuning strategies to lower the graviton mode energy below the continuum. We argue this is a necessary condition enabling the observation of graviton modes and geometric excitations in realistic moir\'e systems.
\end{abstract}

\maketitle

\section{Introduction}

Neutral excitations in two-dimensional (2D) topological phases have been one of the focuses of recent experimental and theoretical studies \cite{Eisenstein2019,Xu2020, Zhang2025,WuReview2024,Wagner2024,kumar2022neutral,khanna2022emergence,balram2024splitting, liu2024resolving,lu2024thermodynamic,Liang2024,long2025spectramagnetorotonchiralgraviton}. They encode both universal topological properties and important dynamical details of the exotic phases of matter \cite{wen1990ground,wen1995topological,Haldane2011,Yang2012, yang2012band,NiuReview2016}. Notable examples of such 2D topological phases include strongly correlated topological bands, which can manifest in both conventional lattices and moir\'e systems, with or without an external magnetic field ~\cite{Klitzing1980,Tsui1982,Wu2021,Xie2021,Sheng2011,Regnault2011,Parameswaran2013,Spanton2018,Huang2021,Zhao2024}. In partially filled topological bands with narrow bandwidths, electron-electron interactions dominate the physics, resulting in ground states with non-trivial topological orders and various emergent excitations. These excitations, arising from the collective motion of electrons, have been extensively studied in the context of fractional quantum Hall (FQH) systems. Among these, Girvin-Macdonald-Platzman (GMP) modes constitute a distinctive branch of low-lying neutral excitations that define the neutral gap of the FQH fluid \cite{Girvin1986,Pinczuk1993,Pinczuk1998}. At long wavelengths, these modes exhibit a quadrupolar structure, transitioning into a dipolar character as the momentum increases \cite{Lee1991,Platzman1996}.

The long-wavelength behavior of GMP modes in FQH is especially intriguing because of the inherent non-commutative quantum geometry of Landau levels (LLs) \cite{Kogan1994}. In this context, the GMP mode represents an area-preserving diffeomorphism within the non-commutative guiding center space \cite{Iso1992,Cappelli1993, flohr1994infinite, Haldane2011, yang2024quantum}. These excitations originate from quantum fluctuations of the \emph{many-body} metric emerging from the strong interactions and are therefore referred to as ``graviton modes (GMs)'' in the FQH literature \cite{Haldane2011, Golkar2016}. The LLs are thus a fascinating platform for studying a non-relativistic quantum gravity in a two-dimensional space \cite{Gromov2017,Nguyen2023}. The existence of emergent GMs and their chiralities has been experimentally verified in Abelian fractional quantum Hall states using polarized Raman scattering ~\cite{Liang2024}, directly confirming their universal properties~\cite{Golkar2016,Nguyen2014,Yuzhu2023,Nguyen2022}. Moreover, certain FQH phases with more intricate Hilbert space structures can give rise to multiple GMs, each corresponding to fluctuations of different conformal Hilbert space metrics \cite{Nguyen2021, Balram2022, Nguyen2022, Yuzhu2023}. Signatures of multi-graviton modes have been predicted at specific filling fractions, such as $\nu = 2/7$ and $\nu = 2/9$, providing deeper insights into the geometric degrees of freedom in FQH phases.

\begin{figure}
\begin{center}
\includegraphics[width=\linewidth]{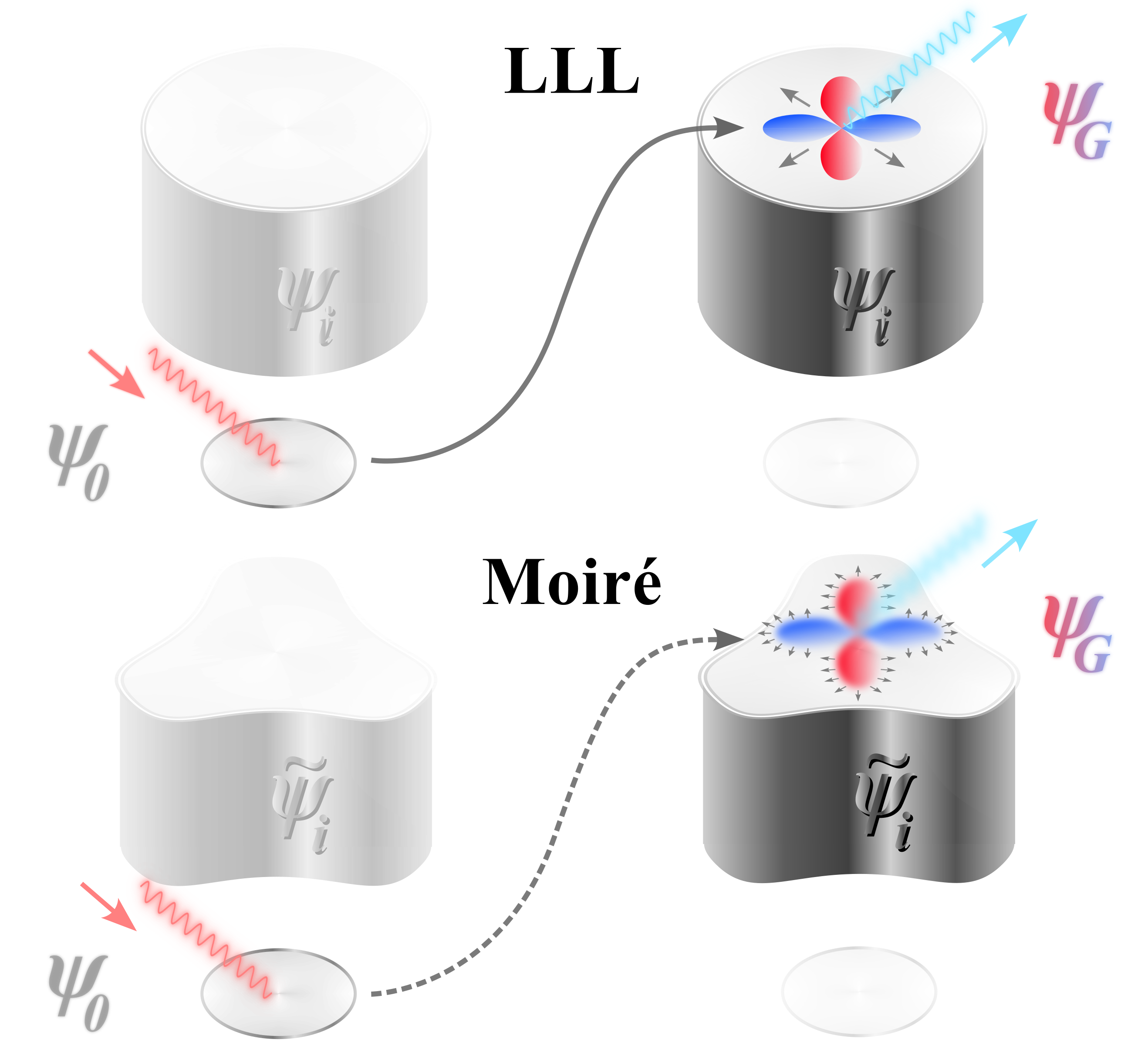}
\caption{\textbf{Schematic illustrating the shorter lifetime of graviton modes (GMs) in moir\'e Chern bands compared to Landau levels (LLs).} In both systems, the GMs $\psi_G$ are excited by incoming polarized beams (red) from the same topological ground state $\psi_0$. These modes possess well-defined ``spins'' due to an emergent guiding-center rotational symmetry. In LLs, all gapped excitations $\psi_i$ maintain continuous rotational symmetry, ensuring that spin-2 graviton modes can only scatter into sectors with the same spin, depicted by the quadrupole structure in the figure. In contrast, moir\'e Chern bands lack continuous rotational symmetry due to the underlying lattice structure. Changing from the LLs to moir\'e Chern bands can be considered as adding perturbations that have little effect on the ground state and GM, but completely reorganize the excited states from $\{\psi_i\}$ to $\{\tilde{\psi}_i\}$. 
Consequently, the highly dense excitations $\tilde{\psi}_i$ do not possess well-defined spins, leading to graviton modes scattering into multiple channels. As a result, in moir\'e systems, there are no resonance peaks from the outgoing polarized beam (blue), implying a significantly short GM lifetime, whereas in LLs, a sharp spectral peak can be observed when scanning different frequencies.} 
\label{Fig_schematic}
\end{center}
\end{figure}

Recent advancements in experiments on the quantum anomalous Hall effect (QAHE) and fractional Chern insulators (FCIs) have shown that FQH-like phenomena can emerge in lattice systems with topologically nontrivial flat bands, even in the absence of an external magnetic field \cite{Park2023, Xu2023, Cai2023, Zeng2023, Redekop2024, Ji2024}. This significant breakthrough has been corroborated by experimental observations of robust fractional Hall conductance \cite{Park2023, Xu2023}, local thermodynamic gap \cite{Redekop2024}, and the imaging of edge states \cite{Ji2024}. In addition, there is significant numerical evidence confirming that essential features from FQH phases are present, including topological degeneracy, entanglement spectra, and quasiparticle statistics \cite{Tang2011, sun2011nearly, neupert2011fractional, Sheng2011, Regnault2011, wang2011fractional, LIU2024515}. The LLs are the simplest topological bands, with zero bandwidth and a uniform quantum geometric tensor, while in 2D quantum materials hosting FCI phases, the band curvature is nonvanishing and the quantum geometry is nonuniform. It is thus natural to ask whether GMs can also emerge in these more intricate topological bands. Moreover, these systems provide an opportunity to explore physics beyond the conventional LL paradigm. The richer quantum geometry and reduced symmetry of the FCIs suggest that the behavior of GMs in these systems could differ significantly, potentially revealing new insights into the interplay of topology, geometry, and dynamics in fractionalized topological phases.

In this paper, we explore the graviton-like excitations in the moir\'e systems such as twisted bilayer graphene (TBG), where FCI phases have been observed \cite{Xie2021}. We extract the essential physics using a simplified ideal flat band (IFB) model, revealing that GMs are significantly weaker compared to the FQH case \cite{Roy2014,Claassen2015, Mera2021, Wang2021,estienne2023ifb,Ledwith2023}. Our numerical results indicate a rapid decay of the GM lifetime as the density of states (DOS) increases, suggesting that the observed behavior may be strongly influenced by finite-size effects. In particular, with Coulomb interactions, a peak in the spectral function emerges in small systems just above the continuum boundary. However, in the thermodynamic limit where the DOS diverges in the continuum, this resonance peak may disappear. Additionally, we show analytically that the suppression of GM lifetimes is due to the strong scattering between different angular momentum sectors, despite the surprising robustness of guiding center rotational invariance of the ground states, as illustrated in Fig.~\ref{Fig_schematic}. Both our analytic and numerical results should be applicable for generic Chern bands, and we suggest a potential approach to stabilize GMs in experimental settings in such systems.

This paper is organized as follows: In Section \ref{sec:ii}, we introduce the IFBs and geometric excitations in FQH/FCI phases. In Section \ref{sec:iii}, we focus on the chiral limit of TBG (cTBG), which forms IFBs that closely resemble LLs, and show the \textit{significantly weaker} spectral peaks of the GMs at filling $\nu=1/3$ in IFBs than in the LLs. In Section \ref{sec:iv}, we examine the DOS to quantitatively confirm such differences in geometric excitations in LLs and IFBs. In Section \ref{sec:v}, we propose a perturbative model Hamiltonian to explain the sensitivity of the GMs to perturbations. We further investigate these phenomena in more realistic models of MoTe$_2$ in Section \ref{sec:vi}, and propose an interaction that can significantly enhance the GM lifetime, guiding potential experimental probes of GMs in FCI phases.

\section{Moir\'e ideal flat bands and geometric excitations\label{sec:ii}}

The absence of a strong magnetic field in FCIs necessitates the fine-tuning of band properties, including bandwidth, topological invariants, and quantum geometry, to support the emergence of FCI phases \cite{Parameswaran2012, Murthy2012,Parameswaran2013,Dobardzic2013, Roy2014}. While toy models based on checkerboard \cite{sun2011nearly, Sheng2011, neupert2011fractional}, ruby \cite{Hu2011}, and kagome \cite{Tang2011} lattices have been proposed to achieve such conditions, an appropriate experimental platform to realize FCIs remained elusive until the discovery of moir\'e systems, such as TBG \cite{Xie2021} and twisted transition metal dichalcogenides (\textit{t}TMDs) \cite{Xu2023,Park2023,Cai2023,Zeng2023}.

We aim to examine the effect of quantum geometry and lattice symmetry on the GMs generally, but are particularly interested in the case of TBG. In the Bistritzer-MacDonald model, the interlayer coupling is described by two parameters, $w_{AA}$ and $w_{AB}$, which give the interlayer tunneling in the AA and AB regions respectively \cite{Bistritzer2011, Nick2020Ground, ledwith2020fractional}. When $w_{AA}$ is neglected, TBG enters the so-called \textit{chiral limit}, where the band at charge-neutrality becomes perfectly flat at the magic angles, with Chern number $\mathcal{C}=1$. While this chiral limit is not entirely realistic (the ratio $w_{AA}/w_{AB}$ is approximately 0.7-0.8 \cite{Koshino2018, Carr2019}), it serves as a good starting point for examining the effect of nonuniform quantum geometry on the GMs. In addition, it has been suggested that the ground state wavefunctions for the realistic case are adiabatically connected to the chiral limit \cite{ledwith2020fractional}, in which case we expect our general conclusions to hold in experimental settings. We will further explain that the general results obtained from the chiral limit hold beyond this idealized case in Section \ref{sec:v}.

In the chiral limit, the TBG band also becomes an IFB. This type of band is the simplest non-trivial generalization of the LLs, where the fluctuation of the Berry curvature $\Omega_{\boldsymbol{k}}$ and the quantum metric $g_{\boldsymbol{k}}^{a b}$ are ``in sync" \cite{Claassen2015, Mera2021, Wang2021}, with the latter becoming positive definite. This relationship is captured by \cite{Roy2014, Claassen2015, Mera2021, Wang2021}:
\begin{equation}
g_{\boldsymbol{k}}^{a b}=\frac{1}{2} \omega^{a b} \Omega_{\boldsymbol{k}},
\label{ideal_band}
\end{equation}
where $\omega^{a b}$ is a positive symmetric matrix with $|\omega^{a b}| = 1$. In experimental systems without a magnetic field, particularly twisted MoTe$_2$ (\textit{t}MoTe$_{2}$), the Chern bands hosting FCI phases are nearly ideal but do not exactly satisfy Eq.~\ref{ideal_band} \cite{ledwith2020fractional, Dong2023CF}.

The LLs represent a special case of IFBs, where the quantum geometry is independent of $\boldsymbol{k}$ and thus uniform, and the only difference between a generic IFB and a LL is the form factor. Under these conditions, an exact mapping can be established between the wavefunctions $\psi$ of LLs to the wavefunctions $\tilde{\psi}$ of IFBs, provided the modulated quantum geometry is properly accounted for. This mapping is given by \cite{Wang2021}:
\begin{equation}
\tilde{\psi}_{\boldsymbol{k}} (\boldsymbol{r})=\mathcal{N}_{\boldsymbol{k}} \cdot \mathcal{B}(\boldsymbol{r})  \cdot \psi_{\boldsymbol{k}} (\boldsymbol{r}), \label{Eq_FCI_wavefunction}
\end{equation}
where $\mathcal{N}_{\boldsymbol{k}}$ is a normalization factor and $\mathcal{B}(\boldsymbol{r})$ is quasi-periodic over the unit cell. These functions encode the quantum geometry fluctuations of the band, with the Berry curvature given by \cite{Wang2021}
\begin{equation}
    \Omega_{\boldsymbol{k}}= -1 + \Delta_{\boldsymbol{k}} \ln \mathcal{N}_{\boldsymbol{k}},
\end{equation}
where $\Delta_{\boldsymbol{k}}$ is the Laplace operator in momentum space. 

$\mathcal{B}(\boldsymbol{r})$ introduces a lattice structure (defined by primitive reciprocal lattice vectors $\boldsymbol{b}$) into the wavefunctions so that it breaks the continuous translational symmetry in the many-body ground states. This can be codified by writing
\begin{equation}
    |\mathcal{B}(\boldsymbol{r})|^2 = \sum_{\boldsymbol{b} \in \text{RLV}} w_{\boldsymbol{b}} e^{i \boldsymbol{b} \cdot \boldsymbol{r}},
    \label{B_r}
\end{equation}
where RLV is the set of all reciprocal lattice vectors. The Fourier coefficients $w_{\boldsymbol{b}}$ then modify the form factor of the band, giving \cite{Wang2021}
\begin{equation}
\begin{aligned}
\mathcal{F}(\boldsymbol{k}_1, \boldsymbol{k}_2) :&= \braket{ \boldsymbol{k}_1| e^{i (\boldsymbol{k}_1 -\boldsymbol{k}_2) \cdot \boldsymbol{r}}|\boldsymbol{k}_2} \notag \\
&= \mathcal{N}_{\boldsymbol{k}_1} \mathcal{N}_{\boldsymbol{k}_2} \sum_{\boldsymbol{b}} w_{\boldsymbol{b}} f^{ \boldsymbol{k}_1, \boldsymbol{k}_2}_{- \boldsymbol{b}},
\end{aligned}
\end{equation}
where $f^{ \boldsymbol{k}_1 \boldsymbol{k}_2}_{- \boldsymbol{b}}$ is the form factor in the LLL:
\begin{equation}
f^{ \boldsymbol{k}_1, \boldsymbol{k}_2}_{- \boldsymbol{b}} = \braket{ \boldsymbol{k}_1| e^{i (\boldsymbol{k}_1 -\boldsymbol{k}_2+ \boldsymbol{b}) \cdot \boldsymbol{r}}|\boldsymbol{k}_2}_{\text{LLL}}.
\end{equation}
As a result, the interaction Hamiltonian in the second quantized form, when projected to the IFB, is given by
\begin{equation}
\begin{aligned}
H_{\text{int}}&=\sum_{\boldsymbol{k}_1, \boldsymbol{k}_2 \in \text{ BZ}}  \sum_{\boldsymbol{q}} V( \boldsymbol{q}) \mathcal{F}(\boldsymbol{k}_1 , \boldsymbol{k}_1+ \boldsymbol{q})  \mathcal{F}(\boldsymbol{k}_2 , \boldsymbol{k}_2- \boldsymbol{q}) \notag \\
& \hspace{3cm} c_{\boldsymbol{k}_1}^{\dagger}c_{\boldsymbol{k}_2 }^{\dagger} c_{ \boldsymbol{k}_2 - \boldsymbol{q}} c_{ \boldsymbol{k}_1+ \boldsymbol{q}} \notag \\
&= \sum_{\boldsymbol{q}} V( \boldsymbol{q})  \sum_{\boldsymbol{k}_1, \boldsymbol{k}_2 \in \text{ BZ}} \sum_{\boldsymbol{b}_1, \boldsymbol{b}_2}  \mathcal{N}_{\boldsymbol{k}_1} \mathcal{N}_{\boldsymbol{k}_2} \mathcal{N}_{\boldsymbol{k}_1+ \boldsymbol{q}} \mathcal{N}_{\boldsymbol{k}_2- \boldsymbol{q}}   
  \notag  \\
& \hspace{1cm} w_{\boldsymbol{b}_1} w_{\boldsymbol{b}_2}  f^{\boldsymbol{k}_1, \boldsymbol{k}_1+\boldsymbol{q}}_{- \boldsymbol{b}_1} f^{\boldsymbol{k}_2, \boldsymbol{k}_2-\boldsymbol{q}}_{- \boldsymbol{b}_2} c_{\boldsymbol{k}_1}^{\dagger}c_{\boldsymbol{k}_2 }^{\dagger} c_{ \boldsymbol{k}_2 - \boldsymbol{q}} c_{\boldsymbol{k}_1+ \boldsymbol{q}} 
\label{cTBG_Ham}.
\end{aligned}
\end{equation}

This formalism allows the interaction physics in IFB to be described in terms of the lowest Landau level (LLL), except with a modified interaction that allows Umklapp processes. For IFBs, it has been established that the leading Umklapp processes dominate, so the relevant quantum geometry of the system can be characterized by only two parameters \cite{Wang2021}. $w_0$ describes the uniform part of the geometry, while $w_1$ gives the Fourier component for each of the smallest non-zero reciprocal lattice vectors, which we take to have the same weight. This means that setting $w_1=0$ results in uniform quantum geometry, reducing the system to the LLL on torus geometry. On the other hand, for the IFB in cTBG, continuum model calculations indicate that the ratio is $w_1/w_0 \sim 0.24$ \cite{Wang2021}. By modifying the parameter $w_1$, we can tune the quantum geometry of the system, offering a minimal model for comparing the GMs between LLs and moir\'e systems.

To calculate the response of GMs, the ground-state metric can be deformed in two ways: The first approach employs a one-body operator to construct microscopic trial wavefunctions within the single-mode approximation (SMA) \cite{Girvin1986}:
\begin{equation}
\left|\psi_{\boldsymbol{q}}\right\rangle=\lim _{\boldsymbol{q} \rightarrow 0} \frac{1}{\sqrt{S_{\boldsymbol{q}}}}\delta \hat{\bar{\rho}}_{\boldsymbol{q}}\left|\psi_0\right\rangle,
\end{equation}
because for translationally invariant ground states, the leading contribution of $\delta \hat{\bar{\rho}}_{\boldsymbol{q}\rightarrow 0}$ acts as the generator of area-preserving deformations \cite{Iso1992,Cappelli1993, flohr1994infinite,Haldane2011,Du2022,kousa2025theorymagnetorotonbandsmoire,paul2025shininglightcollectivemodes}. Here $S_{\boldsymbol{q}}$ is the projected static structure factor and $\delta \hat{\bar{\rho}}_{\boldsymbol{q}} = \hat{\bar{\rho}}_{\boldsymbol{q}} - \left\langle\psi_0\right| \hat{\bar{\rho}}_{\boldsymbol{q}}\left|\psi_0\right\rangle$ is the regularized density operator, both projected to a single Chern band. In particular, for LLs $\hat{\bar{\rho}}_{\boldsymbol{q}}$ represents the guiding center density operator. On compact manifolds such as spheres or tori, however, the momentum $\boldsymbol{q}$ becomes a discrete quantum number, and accessing the long-wavelength limit requires taking the thermodynamic limit.

The second approach to perturbing the metric involves a two-body chiral graviton operator \cite{Yang2016, Liu2018, Liou2019,Nguyen2022}, which corresponds to the spin-$\pm 2$ components of the kinetic part of the LLL stress tensor \cite{Nguyen2021a,Nguyen2022}:
\begin{equation}
\hat{O}_{\pm} = \sum_{\boldsymbol{q}} \left(q_x \pm i q_y\right)^2 V({\boldsymbol{q}})  \hat{\bar{\rho}}_{\boldsymbol{q}} \hat{\bar{\rho}}_{-\boldsymbol{q}},
\label{Opm}
\end{equation}
where $\left(q_x \pm i q_y\right)^2$ represents a chiral d-wave symmetry (the magnetic length $\ell_B\equiv 1$) and the LLL form factor $e^{- |\boldsymbol{q}|^2/2}$ has been absorbed into $V({\boldsymbol{q}})$. The chiral graviton operator $\hat{O}_{\pm}$ and the SMA operator $\delta \hat{\bar{\rho}}_{\boldsymbol{q}\rightarrow 0}$ in gapped FQH states are related to each other through the LLL Ward identities \footnote{The detailed discussion was given in the Supplement Material of Ref \cite{Nguyen2022}}. These two approaches differ in their universality and sensitivity to interaction details: The SMA operator is universal, as it captures neutral excitations arising from density modulations of the ground state. 
In contrast, chiral graviton operators are influenced by specific interaction details and characterize global metric deformations. Both the SMA operator and the chiral graviton operators are experimentally accessible through Raman scattering \cite{Pinczuk1993, Kang2000, Golkar2016,Nguyen2021, Nguyen2021a, Liang2024}, and their measured spectral functions are related to each other by sum rules \cite{Golkar2016,Nguyen2014,Nguyen2021}. Chiral graviton operators can also be directly realized by tilting the magnetic field \cite{papic2013tilt, yang2017anisotropic,Liu2018,yang2024quantum} or applying acoustic waves \cite{Yang2016}. Both operators can be extended to IFBs by appropriately adjusting the form factors to account for the discrete translational symmetry of the Hamiltonian and the nonuniform quantum geometry.

To predict the coupling strength of chiral GM in circularly polarized Raman scattering experiments, one can calculate the spectral function \cite{Nguyen2021a, Liang2024}:
\begin{equation}
I(E) = \sum_n \left|\langle n | \hat{G} | 0 \rangle \right|^2 \delta\left(E - E_n + E_0\right),
\label{spec_func}
\end{equation}
where $E_0$ is the ground state energy and $\hat{G}$ can be either $\delta \hat{\bar{\rho}}_{\boldsymbol{q}\rightarrow 0}$ or $\hat{O}_\sigma$. In IFBs (including LLs), the energy and lifetime of GMs can be extracted from $I(E)$: The position of the resonance peak provides an estimate of the GM energy, while the width of the resonance peak reflects the lifetime of the particle or state. A broader peak indicates a shorter lifetime, consistent with the intrinsic limitations imposed by the uncertainty principle.

\section{Graviton lifetime in moir\'e IFBs \label{sec:iii}}

\begin{figure}
\includegraphics[width=0.95\linewidth]{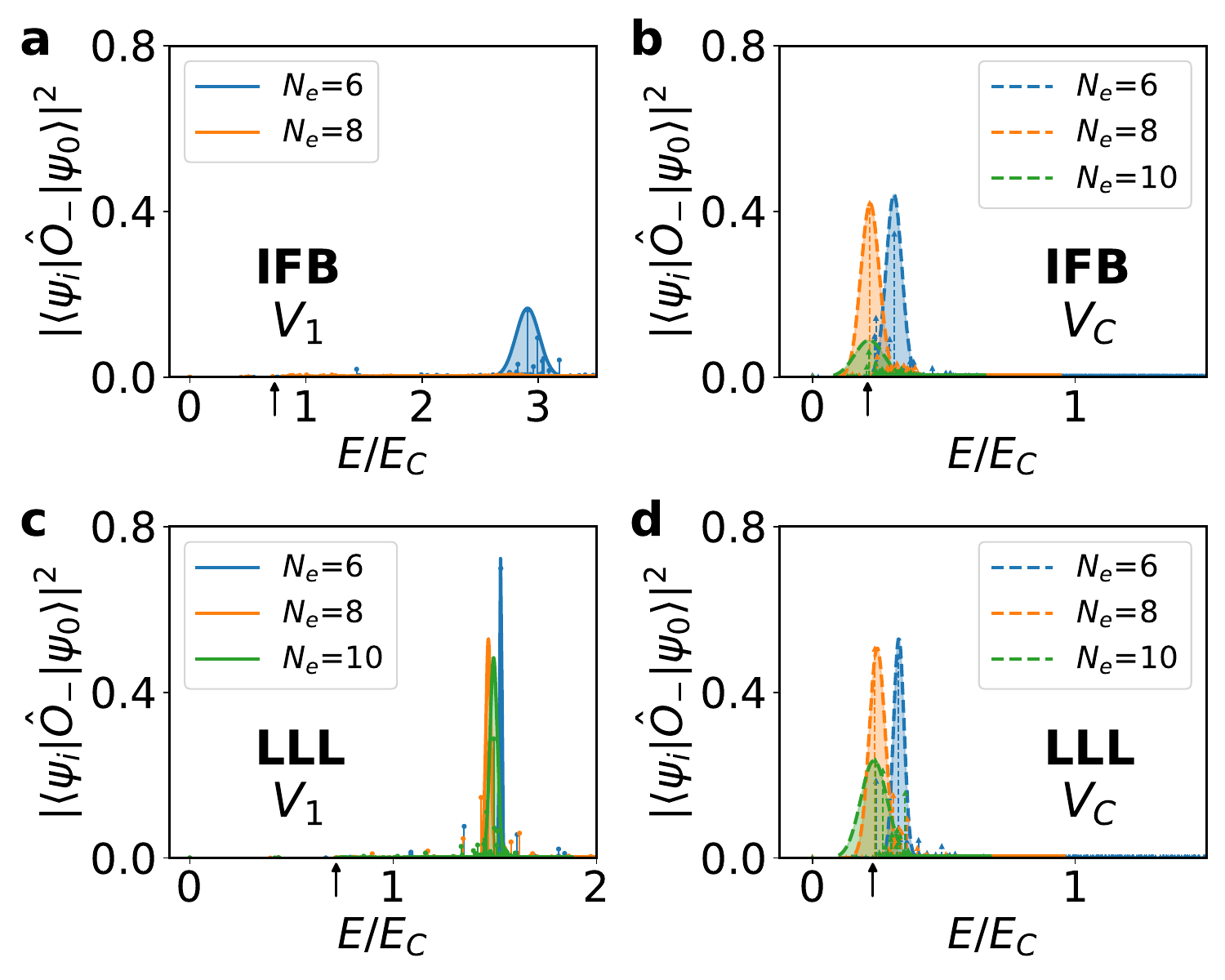}
\caption{\textbf{Spectral functions of the chiral graviton operator at the filling $\nu = 1/3$ in moir\'e IFB and the LLL}. 
All the spectral peaks shown in this paper have been normalized. The system sizes are distinguished by different colors. Arrows denote the boundary positions of the excitation continuum. The energy scale is determined by the Coulomb energy $E_C= e^2/(4 \pi \epsilon l)$, where the characteristic length scale $l = \ell_B$ for LLL and $l = \sqrt{\sqrt{3} / 4 \pi} a_M$ for IFB. Here $a_M$ is the moir\'e lattice constant. 
(\textbf{a}, \textbf{b}) In the IFB of cTBG, a sharp peak persists under $V_C$ even as the system size increases. However, with the $V_1$ pseudopotential, the energy doubles compared to the LLL and dives deeper into the continuum, while the peaks become significantly suppressed as the system size slightly grows. 
(\textbf{c}, \textbf{d}) In the LLL, clear and sharp peaks are observed across all system sizes for both the Coulomb interaction $V_C$ (dashed stems) and the $V_1$ pseudopotential (solid stems). The model Hamiltonian exhibits a particularly pronounced peak, consistent with previous studies on the LLL in torus geometry. These indicate a much shorter GM lifetime in moir\'e IFBs compared to LLs. A Gaussian fit is applied to highlight the GM peak in the plots visually.}
\label{Fig_LLL_SF}
\end{figure}

Building on the formalism presented in the previous section, we numerically compute the GM spectral functions for various system parameters and interactions. These results enable us to predict the signal strength in polarized Raman scattering experiments for GM detection in moir\'e systems \cite{Nguyen2021a}. By comparing the LLL and the moir\'e IFBs, we find that the GM lifetime of the latter becomes significantly shorter under model Hamiltonians. For simplicity we use the parameters of the IFB from cTBG and only keep the leading Umklapp process \cite{Wang2021}. However, all our numerical results and the analytic proof later apply to generic IFB with an arbitrary quasiperiodic $\mathcal B(r)$ in Eq.~\ref{B_r}. 

Focusing on the $1/3$ filling, where the ground states are Laughlin states with three-fold degeneracy, we analyze the GM spectral functions using two types of interactions: the Coulomb interaction $V_{C}({\boldsymbol{q}}) \sim 1/|\boldsymbol{q}|$ and the Haldane pseudopotential $V_1({\boldsymbol{q}}) \sim \mathcal{L}_{1}(|\boldsymbol{q}|^2)$ (where $\mathcal{L}_m(x)$ denotes the Laguerre polynomials). It is important to note that for both the LLL and moir\'e IFB, the model $V_1({\boldsymbol{q}})$ Hamiltonian gives the optimal topological phase with the largest incompressibility gap and exact ground state degeneracy even for finite systems \cite{haldane1983hierarchy,kivelson1985exact,Haldane1985,wen1990ground,Wang2021}. The computed spectral functions for the LLL and moir\'e IFBs are presented in Fig.~\ref{Fig_LLL_SF}, where numerical results confirm that GM peaks with positive chirality are suppressed in all scenarios (so only the negative chirality results are shown), resulting in strong selection rules for Raman channels.

\begin{figure}
\begin{center}
\includegraphics[width=0.97\linewidth]{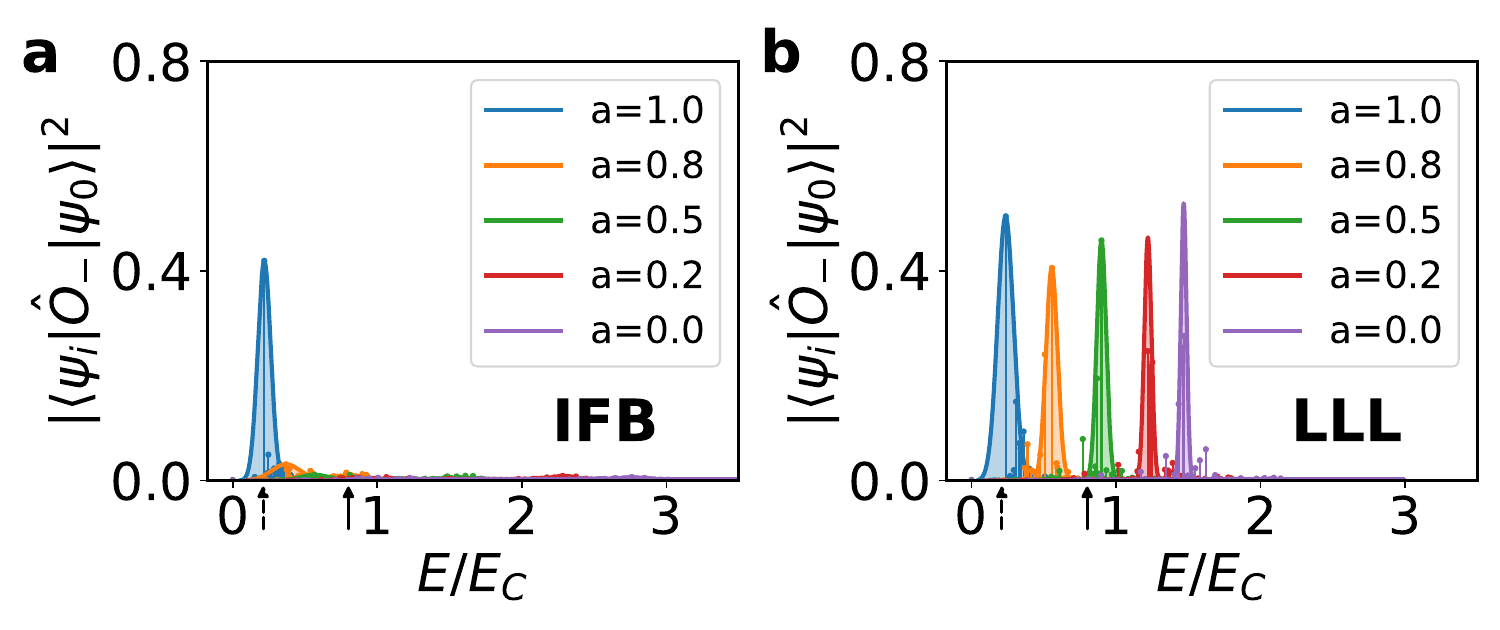}
\caption{\textbf{Spectral function of the toy Hamiltonian $V_a$ at $\nu = 1/3$ with $N_e=8$}. (\textbf{a}) In moir\'e IFBs, reducing the range of the interaction (reducing $a$) significantly suppresses the peak magnitude. (\textbf{b}) However, in the LLL, increasing the range of the interaction does not result in significant changes to the highest peak magnitude. }
\label{Fig_Toy_model_SF}
\end{center}
\end{figure}

For the Coulomb interaction, the peaks in moir\'e IFB (dashed stems in Fig.~\ref{Fig_LLL_SF}b) are only slightly lower than those in the LLL (dashed stems in Fig.~\ref{Fig_LLL_SF}a) for the same system sizes. However, for $V_1$ in IFBs, increasing the system size strongly suppresses the peaks, as shown in Fig.~\ref{Fig_LLL_SF}b. To further investigate this behavior, we employ the following toy model:
\begin{equation}
V_{a}({\boldsymbol{q}}) = a \cdot V_C({\boldsymbol{q}})+(1-a) \cdot V_1({\boldsymbol{q}}), \quad a\in [0,1],
\end{equation}
to compare the effects in both systems. The results, presented in Fig.~\ref{Fig_Toy_model_SF}, reveal a striking contrast between different systems. In the LLL, the GMs under both the $V_1$ and Coulomb interactions are highly robust, despite significant differences in their respective eigenstates. In moir\'e IFBs, however, the behavior is different. While the $V_1$ and Coulomb ground states remain in the same topological phase and show a substantial overlap ($\sim 0.97$ for $N_e=10$), the corresponding GMs (overlap $\sim 0.85$ for $N_e=10$) display distinctly different dynamics. These numerical results on finite systems seem to suggest that, unlike in the LLL, the GM dynamics in moir\'e IFBs are highly sensitive to \textit{specific interactions}, even if such interactions are adiabatically connected leading to the \emph{same topological phase}. Surprisingly, for a model Hamiltonian where the optimal topological phase is realized with an exact ground state, the GM appears to be the weakest.

\section{Finite size effect in moir\'e systems\label{sec:iv}}

A more detailed analysis is needed to understand this unexpected observation that the GM lifetime can be qualitatively different for different interactions within the same topological phase. From the numerical perspective, the lifetime of the GM, or any trial state, is determined both by the nature of the interaction and the number of eigenstates into which it can scatter (which is system size dependent). In the LLL, the optimal interaction for the GMs aligns with the optimal interaction for the topological phase (e.g., $V_1$ for the Laughlin phase at $1/3$): the GMs with $V_1$ exhibit higher peaks, despite a significantly higher DOS near the graviton energy, as compared to the realistic Coulomb interaction \cite{Liou2019, wang2021analytic, Yuzhu2023}. 

In the case of the IFB, this behavior is reversed: GMs with $V_1$ have vanishing lifetimes compared to those under Coulomb interactions, as illustrated in the previous section. However, one should note that the DOS near the graviton energy is \emph{much higher} for the $V_1$ interaction compared to that for the Coulomb interaction. This raises the question: is the weak GM from the $V_1$ interaction simply a result of the large DOS in the numerical computation? If that is the case, the large GM peak with Coulomb interaction is due to the small DOS for the finite system sizes we consider; in the thermodynamic limit when the DOS diverges even for the Coulomb interaction, the GM lifetime may also vanish in contrast to the case in LLs. This implies that the GMs may be very difficult to observe in experiments.

\begin{figure}
\begin{center}
\includegraphics[width=\linewidth]{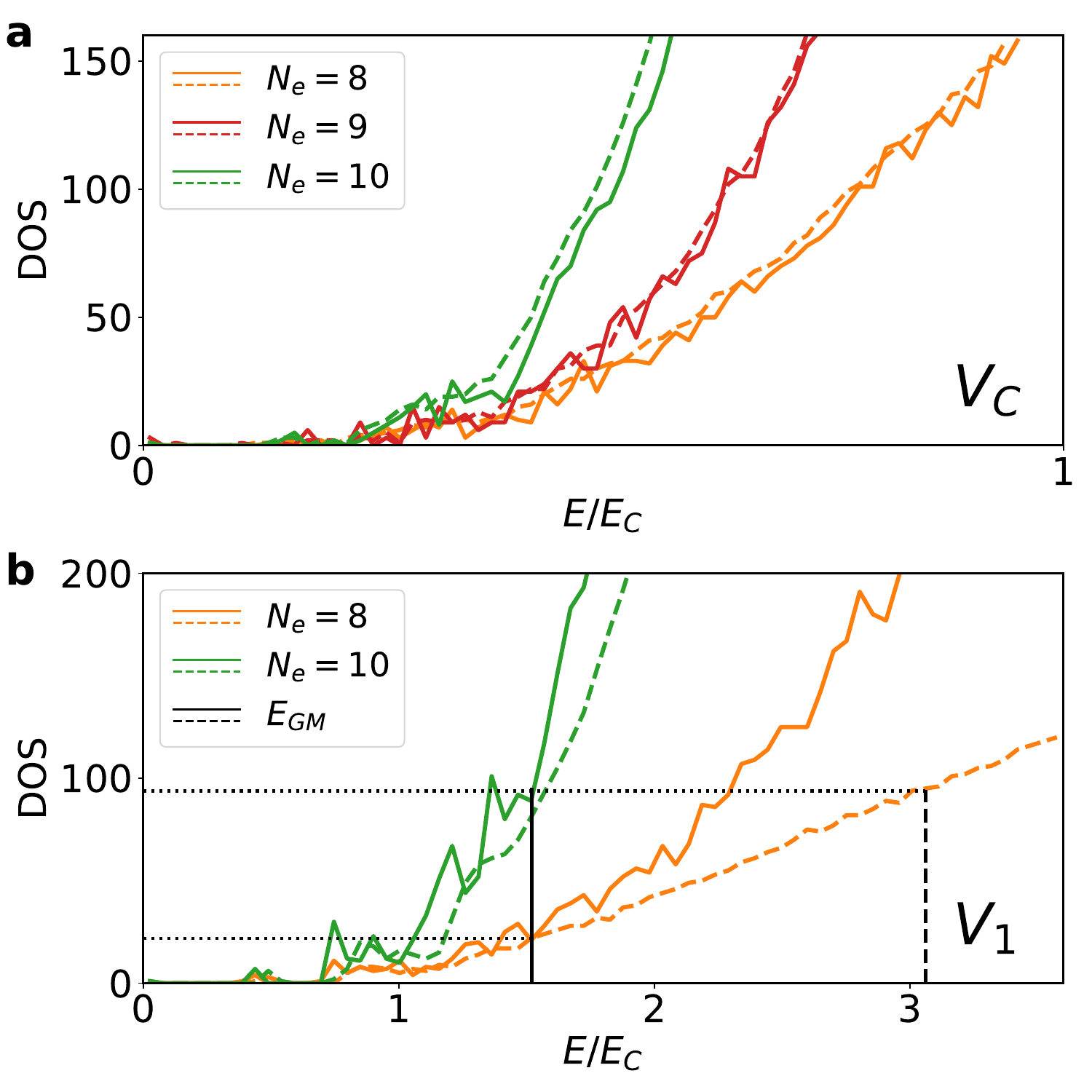}
\caption{\textbf{DOS plots with different settings}. The DOS is defined as the number of states within each interval of $(E_{\text{max}} - E_0)/\Delta$, where $E_{\text{max}}$ is the upper bound of $E$-axis, $E_{0}$ is the ground state energy and we choose $\Delta = 70$. The lowest 3000 eigenstates within the ground state sector are considered. Colors denote different system sizes. Solid lines represent the DOS values in the LLL, while dashed lines are for the IFB. (\textbf{a}) and (\textbf{b}) depict the DOS for the Coulomb interaction and the $V_1$ pseudopotential, respectively. The horizontal line indicates the approximate magnitude of the DOS at the GM energy $E_{GM}$. Notably, in panel \textbf{b}, the DOS at the GM energy for the LLL with $N_e = 10$ (where a sharp GM peak can be observed as shown in Fig.~\ref{Fig_LLL_SF}c) is nearly identical to that of the moir\'e IFB with $N_e = 8$.} 
\label{Fig_DOS}
\end{center}
\end{figure}

We now carry out the DOS analysis based on three key observations: first, the energy of the Coulomb and the $V_1$ GM differs at the same system size; second, a higher GM energy or a larger system results in an increased DOS at a given system size or energy respectively (as shown in Fig.~\ref{Fig_DOS}); third, higher DOS correlates with reduced spectral peak intensity (as shown in Fig.~\ref{Fig_LLL_SF}). These trends are consistent across both LLs and moir\'e IFBs. Moreover, all numerical results are from finite systems, where the spectral functions can be particularly sensitive to the DOS. Thus, controlling the DOS is essential to address the incoherent spectral peaks observed for different interactions in moir\'e IFBs.

To investigate the difference between the LLL and IFB, we control the DOS and the interaction type. The orange lines in Fig.~\ref{Fig_DOS} show the DOS for $N_e = 8$, with panels (a) and (b) presenting the DOS for the Coulomb interaction ($V_C$) and $V_1$, respectively. These results indicate that the DOS at the Coulomb GM energy is comparable for the LLL and IFBs of the same system size, whereas the DOS for $V_1$ exhibits significant variation. By increasing the system size to $N_e = 10$, we observe that the DOS in the LLL at the GM energy becomes nearly identical to the DOS in moir\'e IFBs with $N_e = 8$, as illustrated in Fig.~\ref{Fig_DOS}b.

With both the interaction and the DOS effectively matched, one would expect the GM peaks in the spectral function to show similar profiles for the two systems, if the GMs in the LLL and moir\'e IFBs share the same physical nature. However, the spectral function for $N_e = 10$ with the $V_1$ pseudopotential in the LLL still exhibits a remarkably sharp peak, but the corresponding $N_e=8$ IFB peak is almost completely destroyed, as shown in Fig.~\ref{Fig_LLL_SF}. This counterexample strongly suggests that the rapid decay of the GM peak in moir\'e IFBs with $V_1$ at larger system sizes is closely tied to the specific characteristics of the system itself.

Next, we control the DOS and the system to understand the effect of the interaction. On the LLL, it is well established that for the GM of the Laughlin phase at $1/3$, $V_1$ consistently produces the most robust and well defined GM when all other conditions are the same \cite{wang2021analytic,Yuzhu2023,yang2024quantum}. However, the numerical results in the previous section suggest that this may not hold in IFBs, where the significant DOS differences are ignored. To address this, we resolve the DOS discrepancy by comparing moir\'e IFBs with $V_1$ and $V_C$ at different system sizes for getting close DOS values: As shown in Table~\ref{Table1}, the DOS for $ V_C $ with $N_e = 10$ is approximately $ 2/3 $ that of $ V_1 $ with $ N_e = 6 $. Despite this, the maximal spectral peak intensity ($I^{\text{max}}_\sigma$) for $V_C$ is lower, and the peak becomes more spread out (as indicated by the total overlap in a comparable interval around the GM energy) than for $V_1$. This result demonstrates that even in IFBs, $V_1$ produces a more robust GM than $V_C$ once the DOS is controlled. It further suggests that the high GM peaks observed for $ V_C $ in previous numerical results for moir\'e bands arise from the significantly lower DOS in smaller systems, and these peaks will diminish as system sizes increase \cite{shen2024magnetorotons}.

Combining the analysis above, we conclude that in moir\'e IFBs, the GM differs qualitatively—and potentially fundamentally—from those in the LLL, regardless of the interaction, even though the system remains in the same topological phase. The physics of the GM in moir\'e IFBs is still best captured by the $V_1$ interaction in large systems. The corresponding model GM, originating from the $V_1$ ground state, exhibits a notably shorter lifetime compared to its counterpart in the LLL. Additionally, in numerical studies, the GM lifetime in moir\'e IFBs is primarily determined by the DOS unlike in the LLL. In this case, sharp peaks can be observed for both $V_1$ and Coulomb interactions, provided the DOS is small, suggesting that these observed peaks in moir\'e systems are very likely \textit{finite-size effects}. Thus in experiments, unless the GM energy lies below the continuum, where the DOS remains small even in the thermodynamic limit, the GM will be challenging to observe.

\begin{table}[]
\centering
\resizebox{0.48\textwidth}{!}{%
\begin{tabular}{c|c|c|c|c|c|c}
\hline\hline
$\hat{H}$ &
  $N_e$ &
  $E_g$ &
  \begin{tabular}[c]{@{}c@{}}DOS at $E_g$\end{tabular} &
  \begin{tabular}[c]{@{}c@{}}Total Overlap\end{tabular} &$I^{max}_\sigma$ &
  \begin{tabular}[c]{@{}c@{}}DOS at $I^{max}_\sigma$\end{tabular}   \\ \hline\hline
$\hat{V}_1$ & 6  & 3.1913 & 168 & 0.66163 & 0.16 & 153  \\ \hline\hline
            & 6  & 0.3211 & 19  & 0.60647 & 0.42 & 20   \\ \cline{2-7} 
            & 8  & 0.2914 & 51  & 0.30813 & 0.35 & 18   \\ \cline{2-7} 
\multirow{-3}{*}{$\hat{V}_C$} & 10 & 0.2967 & 115 & 0.58172 & 0.12 &31   \\ \hline\hline
\end{tabular}%
}
\caption{\textbf{Comparison of the spectral peaks in moir\'e IFBs with different interactions and system sizes.} $E_g$ denotes the GM energy, and the DOS is given by the number of states within the interval $[E_g - \Delta/2, E_g + \Delta/2]$, where $\Delta$ for $N_e=6$ with $V_1$ is set as $0.6$ (around $E_g/5$), while $dE$ of Coulomb systems is rescaled by the ratio of GM energy, thus around $0.06$. For $N_e=10$ with Coulomb, even if the DOS is still lower than $N_e=6$ with $V_1$, the peak is lower and broader than the latter. Thus in IFBs, for the same DOS, $V_1$ still gives a more well defined GM than Coulomb.}
\label{Table1}
\end{table}

\section{The moir\'e system as perturbed LLs\label{sec:v}}

The weakness of the GMs in IFBs, or the moir\'e systems in general, can be understood analytically using a a minimal model based on cTBG. In this section, we show that there is an emergent continuous rotational symmetry projected into the moir\'e Chern bands for the ground state. This projected rotational symmetry is analogous to the guiding center rotational symmetry in the Landau levels, which is the only rotational symmetry relevant to the dynamics within a single band \cite{haldane1983hierarchy,Haldane2011}. Even for a general Chern band, this rotational symmetry can be very robust if the incompressibility gap is present. The GMs excited from the ground state thus have a well-defined spin. However, the broken continuous rotational symmetry for the \emph{gapped excitations} causes the spin-$2$ GMs to scatter across all angular momentum sectors. The significantly increased scattering channels suppress the GM peaks in the spectral function, making it much more challenging to observe GMs in IFBs, or lattice Chern bands in general.

To demonstrate this, we first consider a simple first-quantization picture for the interaction in the IFB. Given that the dynamical behavior of GMs is entirely encoded within the intraband properties \cite{Haldane2011,Liou2019,Nguyen2022,Yuzhu2023,yang2024quantum}, we ignore the single-particle normalization factor, $\mathcal{N}_{\boldsymbol{k}}$, in Eq.~\ref{Eq_FCI_wavefunction}, to simplify our analytical analysis \footnote{Here, we use intuition from the FQH, where the typical low-energy dynamics within a single LL are largely insensitive to normalization. As a result, the low-energy properties of fractional topological fluids remain consistent across identical topological phases in different LLs, despite variations in single-particle normalization.}. When we do so, the electrons in the IFB, interacting via $V(\boldsymbol{r}_1-\boldsymbol{r}_2)$, can be mapped on to electrons in a LLL, interacting via the effective interaction \cite{Wang2021}:
\begin{equation}
\tilde{V}(\boldsymbol{r}_1, \boldsymbol{r}_2) = |\mathcal{B}(\boldsymbol{r}_1)|^2 |\mathcal{B}(\boldsymbol{r}_2)|^2 V(\boldsymbol{r}_1 - \boldsymbol{r}_2). \label{Eq_effective_interaction_1}
\end{equation}
Because $|\mathcal{B}(\boldsymbol{r})|^2$ is periodic over the unit cell, we see that this effective interaction breaks the continuous translation symmetry in the LLL down to the discrete translational symmetry of the IFB. For simplicity, we retain only the terms linear in $w_1$ in Eq.~\ref{cTBG_Ham} to construct a minimal model to capture the essential physics. In this case, the symmetry breaking part of the interaction can then be written as
\begin{eqnarray}\label{linearized}
 \delta V(\boldsymbol{r}_1, \boldsymbol{r}_2) =2w_1w_0 \sum_{\boldsymbol{b}}   e^{i \boldsymbol{b} \cdot \boldsymbol{R} } \cos(\frac{1}{2} \boldsymbol{b} \cdot \boldsymbol{r}) V(\boldsymbol{r})
\end{eqnarray}
where $\boldsymbol{r}= \boldsymbol{r}_1- \boldsymbol{r}_2$ is the relative coordinate and $\boldsymbol{R}= \frac{\boldsymbol{r}_1+ \boldsymbol{r}_2}{2}$ is the centre-of-mass coordinate, and the summation is over the smallest reciprocal lattice vectors (sRLV). One can prove analytically that, with the shortest range interaction for fermions, $V(\boldsymbol r)=\nabla''\delta\left(\boldsymbol r\right)$, the isotropic Laughlin state at $\nu=1/3$ is the exact zero energy state for the effective LLL model. This is despite the explicit breaking of the continuous rotational symmetry of the Hamiltonian, which is evident in the anisotropic dependence of $\delta V(\boldsymbol{r}_1, \boldsymbol{r}_2)$ on the center-of-mass coordinate $\boldsymbol{R}$. We therefore see that the excited states will generally be affected by scattering that changes the total angular momentum of a pair of electrons, while the ground state remains isotropic.

While this argument shows the emergent rotational symmetry of the ground state, despite the effective interaction breaking the symmetry, it is quite particular to the IFB. By considering the interaction in momentum basis, we can obtain expressions that may be more easily generalizable to other Chern bands. To do this, we again use the expression for the full Hamiltonian given in Eq.~\ref{cTBG_Ham} and ignore the normalization factors. This time, however, we use the relation \cite{Wang2021} 
\begin{equation}
     f^{\boldsymbol{t}, \boldsymbol{t}+ \boldsymbol{q}}_{- \boldsymbol{b}_i} =\eta(\boldsymbol{b}_i)e^{-\frac{1}{4}|  \boldsymbol{b}_i|^2}   e^{-i \boldsymbol{t} \times \boldsymbol{b}_i} e^{\frac{1}{2} b_i^* q}  f^{\boldsymbol{t},   \boldsymbol{t}+\boldsymbol{q}}_{\boldsymbol{0}},
\end{equation}
with the complex variables $q=q_x +i q_y$, $b_i=b_{i,x}+i b_{i,y}$, and where $\eta_{\boldsymbol{b}}=+1$ for $\boldsymbol{b}/2 \in $ RLV and $-1$ otherwise. This allows us to extract all the dependence on the RLV into a generalized momentum-space interaction $\tilde{V}_{\boldsymbol{q}, \boldsymbol{s}, \boldsymbol{t}}$. We can then write the full Hamiltonian $\hat H_s$ on the torus as a perturbation to the Hamiltonian in the LLL as follows:
\begin{equation}
\begin{aligned}
\hat H_s= \sum_{\boldsymbol{q}}\sum_{\boldsymbol{s}, \boldsymbol{t} \in BZ} \tilde{V}_{\boldsymbol{q}, \boldsymbol{s}, \boldsymbol{t}}  f^{\boldsymbol{s},  \boldsymbol{s} -\boldsymbol{q} }_{\boldsymbol{0}} f^{\boldsymbol{t}, \boldsymbol{t} +\boldsymbol{q} }_{\boldsymbol{0}} c^{\dagger}_{\boldsymbol{s}} c^{\dagger}_{\boldsymbol{t}} c_{ \boldsymbol{t} +\boldsymbol{q} } c_{\boldsymbol{s} -\boldsymbol{q} }, 
\label{Eq_LLL_perturbation}
\end{aligned}
\end{equation}
where in $\tilde{V}_{\boldsymbol{q}, \boldsymbol{s}, \boldsymbol{t}} = V(\boldsymbol{q})\left(1+ \varepsilon_{\boldsymbol{q},\boldsymbol{s},\boldsymbol{t}}\right)$, $\boldsymbol{q}$ is the momentum transfer (Here we take $w_0 = 1$ without loss of generality). By replacing the IFB creation and annihilation operators with LLL operators, we can map the IFB system to an LLL problem with the same representation for the Hamiltonian in momentum basis. The broken continuous translational symmetry from the original lattice system is encoded in the perturbation:
\begin{equation}
\begin{aligned}
\varepsilon_{\boldsymbol{q},\boldsymbol{s}, \boldsymbol{t}}
=&\sum_{\boldsymbol{b}_i \neq \boldsymbol{0}} w_{\boldsymbol{b}_i} \left(\mathcal{F}_{\boldsymbol{b}_i,-\boldsymbol{q}, \boldsymbol{s}} +  \mathcal{F}_{\boldsymbol{b}_i,\boldsymbol{q}, \boldsymbol{t}} \right)\\
&+\sum_{\boldsymbol{b}_i,\boldsymbol{b}_j \neq \boldsymbol{0}} w_{\boldsymbol{b}_i}w_{\boldsymbol{b}_j}\mathcal F_{\boldsymbol{b}_i,-\boldsymbol{q}, \boldsymbol{s}}\mathcal F_{\boldsymbol{b}_j,\boldsymbol{q}, \boldsymbol{t}}\label{delta_V},
\end{aligned}
\end{equation}
where $\mathcal{F}_{\boldsymbol{b},\boldsymbol{q},\boldsymbol{t}} \equiv \eta_{\boldsymbol{b}} e^{-\frac{1}{4}\left|\boldsymbol{b}\right|^2} e^{-i \boldsymbol{t} \times \boldsymbol{b}} e^{\frac{1}{2} b^* q}$. In the case where $w_{\boldsymbol{b}_i}=0$, the Hamiltonian is equivalent to the LLL Hamiltonian in the momentum basis.

Interestingly, even beyond the linearized special case in Eq.(\ref{linearized}), the ground state wavefunctions are invariant as exact zero energy Laughlin states with arbitrary choices of $w_{\boldsymbol{b}_i}$ for $V(\boldsymbol{q})=V_1$, with the exact three-fold degeneracy at $\nu=1/3$. This implies that the universal properties of the GM still hold, including the chirality \cite{Liou2019,Nguyen2021a,Yuzhu2023}, which has been confirmed by extensive numerical computations. The universal behavior can be analytically proven by expanding Eq.~\ref{delta_V} in the orthonormal basis of the generalized pseudopotentials (PPs). Crucially, the perturbation for nonzero $w_{\boldsymbol{b}_i}$ is \textit{holomorphic} in $q$, apart from in $V_1(\boldsymbol{q})$, reflecting the chiral nature of the IFB. These holomorphic terms do not compromise the Hermiticity of the Hamiltonian in the torus geometry \footnote{As long as the inversion symmetry in $\boldsymbol{b}_i$ vectors is preserved, the Hamiltonian will remain Hermitian. Refer to the supplementary material for more details.}. Because of this holomorphicity, the expansion in terms of generalized pseudopotentials only involves the $V^{+}_{1,m}$ terms \footnote{Detailed derivations are given in the supplementary material.}: 
\begin{eqnarray}\label{gpp}
V_1(\boldsymbol{q}) \varepsilon_{\boldsymbol{q}, \boldsymbol{s}, \boldsymbol{t}}=\sum_{m} \lambda_{1,m,\boldsymbol{s},\boldsymbol{t}}V^{+}_{1,m}(\boldsymbol{q}), 
\end{eqnarray}
where $V^{+}_{1,m}(\boldsymbol{q})\sim q^m  \mathcal L_1^m\left(|\boldsymbol{q}|^2\right)$ is the generalized PP with $\mathcal L^m_n\left(x\right)$ the generalized Laguerre polynomial 
$\mathcal L^m_n\left(x\right)$  \cite{yang2017generalized, yang2017anisotropic, Wang2021}. Physically, the $V^{+}_{1,m}$ pseudopotentials only energetically penalize a pair of electrons if its relative angular momentum is $ \Delta L = 1$. The Laughlin model state and its quasihole states thus have exact zero energy no matter how complicated $\lambda_{1,m,\boldsymbol{s},\boldsymbol{t}}$ are, as any pair of electrons has at least $\Delta L \geq 3$.

\begin{figure}
\begin{center}
\includegraphics[width=0.98\linewidth]{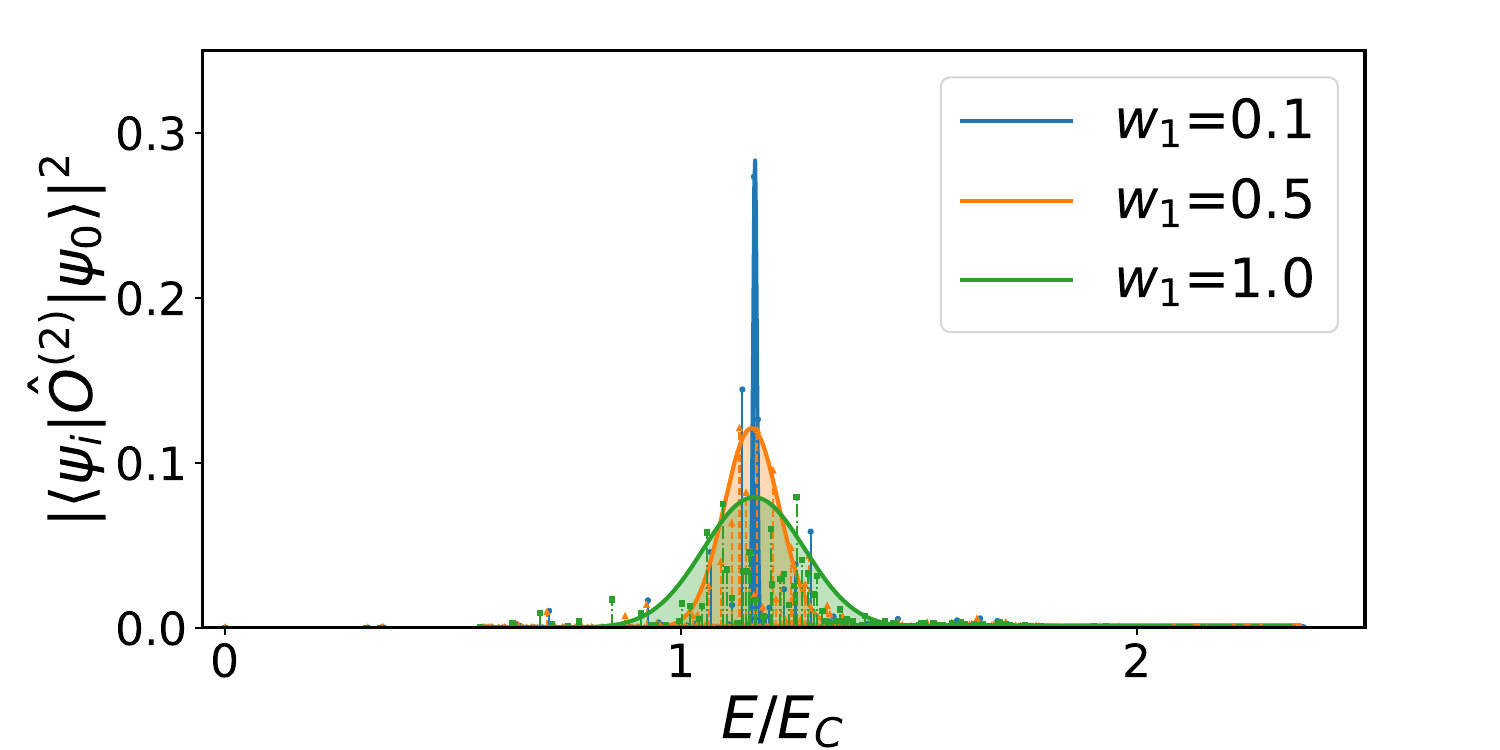}
\caption{\textbf{Spectral function of $\hat{H}_s$ with different perturbation strengths.} The GM peak diminishes rapidly with increasing perturbation strength. The $w_1$ in the legend is relative to the cTBG value. The spectral peaks decrease monotonically with larger $w_1$.}
\label{Fig_toy}
\end{center}
\end{figure}

One can thus understand the vanishing lifetime of the GM from $\hat H_s$ with nonzero $w_{\boldsymbol{b}_i}$ in a rather transparent manner: the many-body GM, from the geometric deformation of the ground state, is invariant with $w_{\boldsymbol{b}_i}$. In the thermodynamic limit, the ground state (and the zero energy quasihole states) exhibits an emergent guiding center rotational symmetry, as it is \emph{identical} to the Laughlin state in the LLL apart from the single particle normalization. This emergent symmetry arises even though the interaction explicitly breaks the continuous rotational symmetry in the moir\'e IFBs. On the contrary, the gapped excitations in the continuum are strongly affected by Eq.(\ref{gpp}) and become highly anisotropic, mixing different angular momentum sectors. Naturally the rotationally invariant GM in such systems will be strongly scattered if it lies within the continuum, in contrast to the cases in the LLs.

The emergent rotational symmetry should also be robust in realistic systems or more general models (e.g., beyond the model interaction or the chiral limit), as long as the topological phase itself is robust with a large incompressible gap compared to the perturbations to the chiral interaction. This is because the perturbations to the chiral interaction are suppressed by the incompressibility gap, rendering the ground state (and thus the GM) only slightly anisotropic. In contrast, $\varepsilon_{\boldsymbol{q}, \boldsymbol{s}, \boldsymbol{t}}$ significantly perturbs the continuous gapped excitations of $\hat H_{LLL}$, scattering the excitations between different angular momentum sectors. In particular, the spin-2 modes present in the LLL spectrum, responsible for the narrow spectral peak, will scatter with other higher spin modes by $\varepsilon_{\boldsymbol{q}, \boldsymbol{s}, \boldsymbol{t}}$. This explanation is numerically confirmed by Fig.~\ref{Fig_toy}, where the GM peak of $\hat{H}_s$ rapidly decreases with increasing $w_1$.

Thus for moir\'e systems realizing Chern bands that well approximate the IFB conditions (e.g., the TBG and MoTe$_2$ systems), the main physics is captured by the dynamical properties of $\hat H_{s}$. There is an explicit breaking of rotational symmetry of the interaction within the flat band, for any interaction supporting a robust topological phase (e.g., Coulomb or $V_1$ interaction). Such interactions can always be expressed as an anisotropic perturbation to $\hat H_{LLL}$, just like Eq.~\ref{Eq_LLL_perturbation} (but with different $\varepsilon_{\boldsymbol{q}, \boldsymbol{s}, \boldsymbol{t}}$ depending on the details of the system). Interestingly, as explained above, the ground state (and thus the GM emerging from the geometric deformation of the ground state) is hardly perturbed by $\varepsilon_{\boldsymbol{q}, \boldsymbol{s}, \boldsymbol{t}}$ due to the presence of the incompressibility gap, as argued above for $\hat H_s$ and also numerically verified for other models. The scattering of the GM by the anisotropic continuum is the fundamental reason why GMs fail to exhibit a long lifetime in moir\'e systems.

It is useful to recall that microscopic Hamiltonians of the moir\'e systems (or lattice systems in general) only have discrete rotational symmetry (e.g., $C_3$ symmetry in TBG), one thus naively would expect the quantum fluid to have only discrete rotational symmetry both below or above the incompressibility gap. Geometric excitations from the ground state thus seem complicated, and not much can be said about their dynamics other than from numerical computations with finite system sizes. The results from numerical computations are challenging to interpret due to the small DOS, as discussed in the previous section. Generically one would argue that the GMs (which may not even be well defined due to discrete symmetry) are naturally weak and so the spectral peaks found in numerics are surprising. The discovery that the ground state (or any states below the gap) and thus the geometric excitation have emergent continuous rotational symmetry, while gapped excitations do not, leads to two main messages: firstly it gives a firm understanding on why GMs in the continuum will have vanishing lifetime despite the numerical results from small systems; secondly the GMs are still well defined and they can in principle be measured when properly tuned, as we will discuss in detail in the next section.

\section{Experimental discussions based on \lowercase{\textit{t}}TMD systems\label{sec:vi}}
We have demonstrated that GMs in moi\'re systems (or periodic lattice systems in general) are intrinsically weak due to the robustness of the isotropic ground state of the topological phases, even if the Hamiltonian explicitly breaks continuous rotational symmetry. As a result, measuring GMs experimentally becomes extremely challenging when the GM energy lies within the continuum. Additionally, caution is required when interpreting numerical results for finite-size systems with Coulomb interactions (e.g., in Ref.~\cite{shen2024magnetorotons}). Thus to detect GMs in experimental systems, it is crucial to tune the GM energy below the continuum. This ensures that, like the ground state, they are protected by an energy gap to the continuum and exhibit an emergent guiding center rotational symmetry. The key principle for tuning the interaction is to reduce its short-range components \cite{yang20nematic,wang2021analytic,Ajit24splitting}. Under these conditions, the DOS at the GM energy no longer diverges with system size, and a numerically observed narrow spectral peak for small system sizes can remain robust in the thermodynamic limit.

\begin{figure}
\begin{center}
\includegraphics[width=0.98\linewidth]{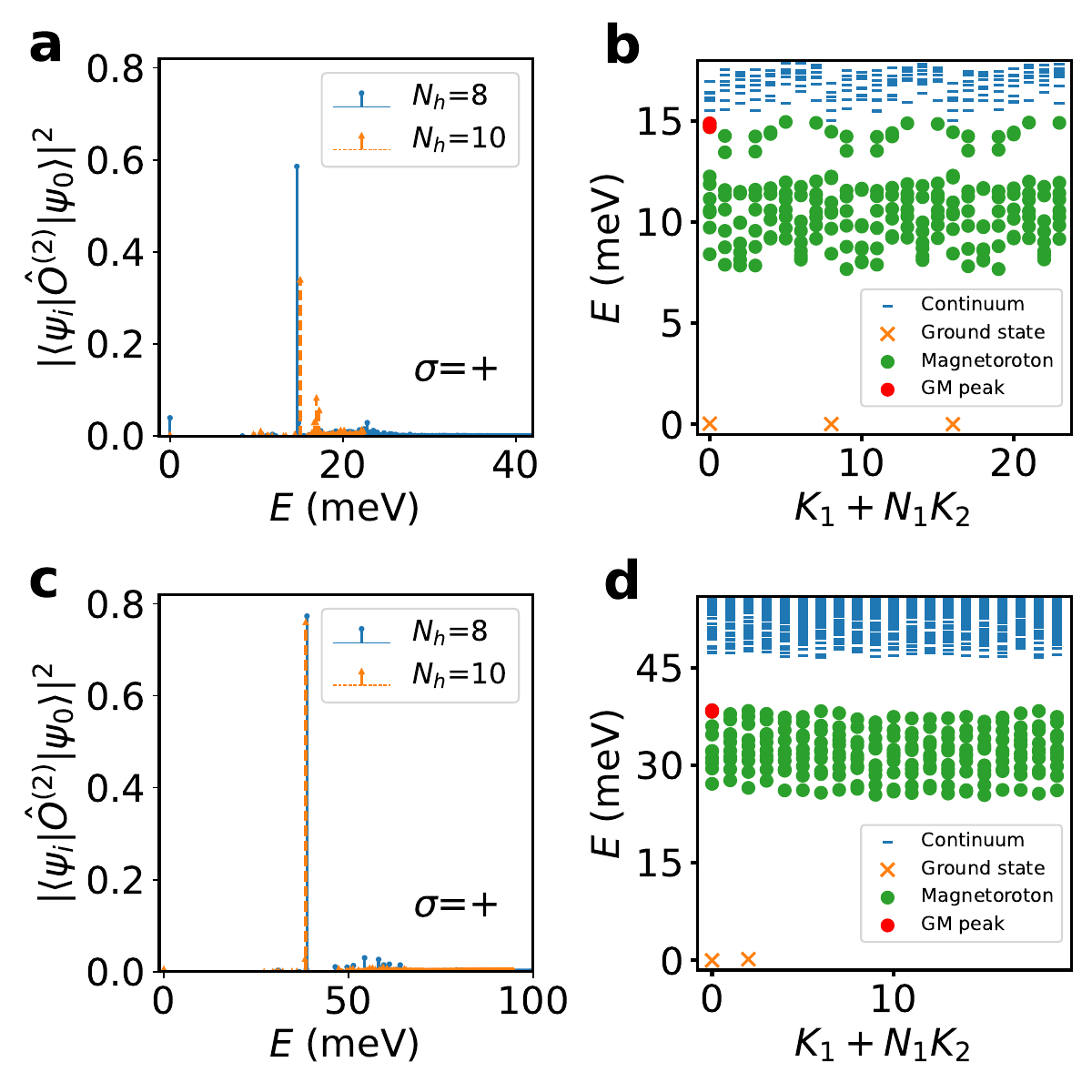}
\caption{\textbf{Spectral functions and spectra of modified Coulomb interaction $V_{\text{ZDS}}$ with effective thickness $\lambda=0.2$ in \textit{t}MoTe$_2$.} For the fermionic system (\textbf{a}, \textbf{b}) with the filling $\nu=-2/3$, the graviton mode (GM) peak is enhanced by approximately 50\% compared to the pure Coulomb case (shown in the supplementary materials). However, the peak strength still decreases as the system size increases. In panel \textbf{b}, the GM position (red dot) lies slightly below the continuum. In contrast, the bosonic system (\textbf{c}, \textbf{d}) with $\nu=-1/2$ exhibits a more robust response. The spectral peak remains strong and even increases slightly with system size. Moreover, the GM position is fully tunable to fall within the magnetoroton modes and becomes well-separated from the continuum. The hole number in \textbf{b} and \textbf{d} is $N_h=10$, where $(K_1, K_2)$ gives the position in the first Brillouin zone with $N_1$ as the number of sites along one of the primitive lattice vectors.}
\label{Fig_MoTe2}
\end{center}
\end{figure}


The analytical results from previous sections are also applicable to various more realistic models beyond the IFB limit, as we have checked extensively with numerics. Here it is useful to focus on one realistic model and calculate the GM spectral functions for \textit{t}MoTe$_2$, where FCI phases have been observed in the absence of magnetic fields \cite{Xu2023,Park2023,Cai2023,Zeng2023}. In Fig.~\ref{Fig_MoTe2}, we present results for a continuum model of \textit{t}MoTe$_2$ with the parameters $(V, \psi, w, \epsilon)$ = ($20.8$meV, $107.7^\circ$, $-23.8$meV, $5$) at the twist angle $3.89^\circ$ and the hole filling $\nu = -2/3$ (so the chirality is opposite to the IFB results above) \cite{wu2019mote2,chong2024mote2}. The chiral graviton operators are taken as the $L=2$ representation of the $D_3$ group to fulfill the periodic boundary conditions and moir\'e lattice symmetry. The GM spectral function shows that \textit{t}MoTe$_2$ exhibits the same behavior as IFBs. Specifically, under pure Coulomb interaction, the GM peak is located near the continuum boundary, with a magnitude of approximately $0.4$. However, with the $V_1$ pseudopotential, the peak shifts deeper into the continuum, reducing in magnitude to around $0.03$. These results are in excellent agreement with those observed for IFBs.

In existing experimental setups, a dual-gate geometry is commonly used to induce hole doping, which introduces a screening effect on the Coulomb interaction. Accordingly, most theoretical calculations consider Coulomb interactions screened by symmetric dual gates \cite{devakul2021magic,reddy23mote2,chong2024mote2}, consistent with experimental configurations \cite{Xu2023,Park2023,Cai2023,Zeng2023}:
\begin{equation}
V_{SC}(\boldsymbol{q}) = \frac{2 \pi e^2}{\epsilon} \frac{\tanh (\xi |\boldsymbol{q}| / 2)}{|\boldsymbol{q}|},
\end{equation}
where $\xi$ represents the screening length, corresponding to the distance to the gate plates. The screening effect suppresses the long-range components, effectively enhancing the short-range components. This would increase the GM energy relative to the continuum. However, its impact is limited since $\xi$ is typically several times larger than the moir\'e lattice constant $a_M$ in common experimental setups \cite{Xu2023,Cai2023}. This means that we expect the GM to have similar strength for the realistic screened interaction as it does for the Coulomb interaction that we considered previously.


To strengthen the GM, we propose a potential interaction mechanism to effectively soften the GM:
\begin{equation}
V_{ZDS}(\boldsymbol{q}) = \frac{2 \pi  e^2}{\epsilon} \cdot \frac{e^{- \lambda \cdot a_M \cdot |\boldsymbol{q}|}}{  |\boldsymbol{q}|}
\end{equation}
Such interaction is the Zhang-Das-Sarma (ZDS) potential, commonly used to describe finite-thickness effects in QH systems \cite{zhang1986excitation,peterson08orbital}. This interaction has smaller short-range components than the standard Coulomb interaction, making it effective in reducing the GM energy. Here, $\lambda \cdot a_M$ represents the ``thickness" of the wavefunctions in \textit{t}MoTe$_2$, accounting for the spread of electronic wavefunctions in the out-of-plane direction. As shown in Fig.~\ref{Fig_MoTe2}b, a slight increase in $\lambda$ shifts the GM peak downward from the continuum boundary and enhances its magnitude compared to the pure Coulomb interaction as confirmed by Fig.~\ref{Fig_MoTe2}a \footnote{The results for pure Coulomb and more $\lambda$ values are shown in the supplementary material}. Experimentally, this increase in the effective thickness of the \textit{t}MoTe$_2$ sample can be achieved by inserting a thin dielectric medium between the twisted layers or using twisted multilayer MoTe$_2$.

One concern arising from Fig.~\ref{Fig_MoTe2}a is that the GM peak continues to decrease with increasing system size because the GM is not significantly away from the continuum, making it uncertain whether the proposed tuning strategy can sustain the GM peak in the thermodynamic limit. Drawing from insights gained in LLs, we propose a more robust approach for probing GMs in moir\'e Chern bands by utilizing bosonic systems. In bosonic FQH phases, the magnetoroton modes tend to be flatter, and the GM modes generally have lower energy, increasing the likelihood of achieving full separation from the continuum \cite{Repellin2014, Liou2019, liu2024resolving}.

Experimentally, bosonic FQH phases can be realized using ultracold atoms in optical lattices \cite{sorensen2005optical, cooper2013reaching, He2017realize, leonard2023realization} or interacting photons \cite{wang2024realization}. If the quantum geometric properties of the moir\'e lattice can be encoded into an experimental system featuring bosonic FCI phases with tunable interactions, this approach will provide significant advantages. Fig.~\ref{Fig_MoTe2}c presents the spectral function for the ZDS interaction with $\lambda = 0.2$ in \textit{t}MoTe$_2$, demonstrating that the GM peak is significantly enhanced compared to the fermionic case, with \textit{minimal decay} as the system size increases. Furthermore, Fig.~\ref{Fig_MoTe2}d confirms that the GM is fully gapped from the continuum and falls within the magnetoroton mode, further supporting the robustness of this approach. Although the realization of bosonic FQH phases is in an early stage, these results provide compelling evidence for the feasibility of probing GMs in bosonic moir\'e Chern bands as experimental techniques continue to advance. However, regardless of particle statistics, it is essential to maintain the effective interlayer distance $ \lambda$ within an optimal range. A significant increase in $ \lambda$ may induce phase transitions that destabilize the FCI phase, which requires further investigation and will be left for future studies. 

\section{Conclusion and outlook}
By analyzing the spectral functions of the chiral graviton operators in moir\'e lattice Chern bands, we find that GMs become significantly weaker than in LLs. Similar behaviors are also seen with GMs generated with SMA in the long wavelength limit. Specifically, the GM lifetime in lattice Chern bands decreases rapidly with an increase in the density of states, a phenomenon further exacerbated when the GM resides within the excitation continuum. This behavior is well captured by a minimal theoretical model, which reveals the mismatch between the symmetries of the ground state and the excited states in generic lattice Chern bands. Moreover, we propose that probing the graviton mode could be achieved by further suppressing the short-range component of the potential to position the GM below the continuum or using a bosonic system that can realize the FCI phases with moir\'e potential. Additionally, we suggest a potential tuning strategy to enhance the GM resonance peak in experimental settings.

Despite this progress, many questions remain regarding the GM and other neutral excitations in moir\'e materials. One key question is whether the graviton mode can be shifted below the continuum in experiments by tuning the effective interactions without inducing a phase transition. Another promising direction is to explore other neutral collective modes in Chern bands, such as finite-momentum GMP modes, gravitino modes, higher-spin modes, and the multiple graviton modes that emerge at different filling fractions in fractional quantum Hall systems. One can also look into the counterparts of higher LLs' neutral excitations in the moir\'e systems. Furthermore, the extremely short lifetime of GMs may indicate a breakdown of the single-mode approximation in the long-wavelength limit of lattice Chern bands. This suggests the possible emergence of novel low-lying neutral excitations in FCI phases, which requires deeper investigation and understanding.

\begin{acknowledgments}
We thank Z. Y. Meng, X. Shen, Z. Liu, L. Du, M. Long, H. Lu, Q. Xu, B. Peng, and Ha Q. Trung for useful discussions. This work is supported by the National Research Foundation, Singapore under the NRF Fellowship Award (NRF-NRFF12-2020-005), Singapore Ministry of Education (MOE) Academic Research Fund Tier 3 Grant (No. MOE-MOET32023-0003) “Quantum Geometric Advantage”, and Singapore Ministry of Education (MOE) Academic Research Fund Tier 2 Grant (No. MOE-T2EP50124-0017). J. H. and Y. B. K. are supported by the Natural Sciences and Engineering Research Council (NSERC) of Canada (NSERC Grant No. RGPIN-2023-03296) and the Center for Quantum Materials at the University of Toronto. D.X.N. is supported by the Institute for
Basic Science in Korea through the Project IBS-R024-D1.
\end{acknowledgments}

\bibliography{QAH_references}{}


\appendix
\counterwithin{figure}{section}

\setcounter{table}{0}
\renewcommand{\thetable}{A.\arabic{table}}
\onecolumngrid  
\newpage

\begin{center}
    {\large \textbf{Supplementary Materials for ``Dynamics and lifetime of geometric excitations in moir\'e systems''}}  
\end{center}

\section{Extended numerical analysis of ideal flat bands} 
\label{sec:supp_A}

In this section, we will present more detailed numerical results, including the spectra of different interactions at the $1/3$ filling in the Landau levels (LLs) and ideal flat bands (IFBs) of chiral twisted bilayer graphene (cTBG), the normalization factor of chiral graviton operators, the spectral functions without normalization, the spectral functions in IFBs with different $w_1$, and the overlap between the ground states/graviton modes in different systems. Note that we always use $w_1=0$ to denote the IFBs with uniform quantum geometry (i.e., the lowest LL) and $w_1=1$ for the IFBs in cTBG. 

\begin{figure}[p]
\includegraphics[width=0.66\linewidth]{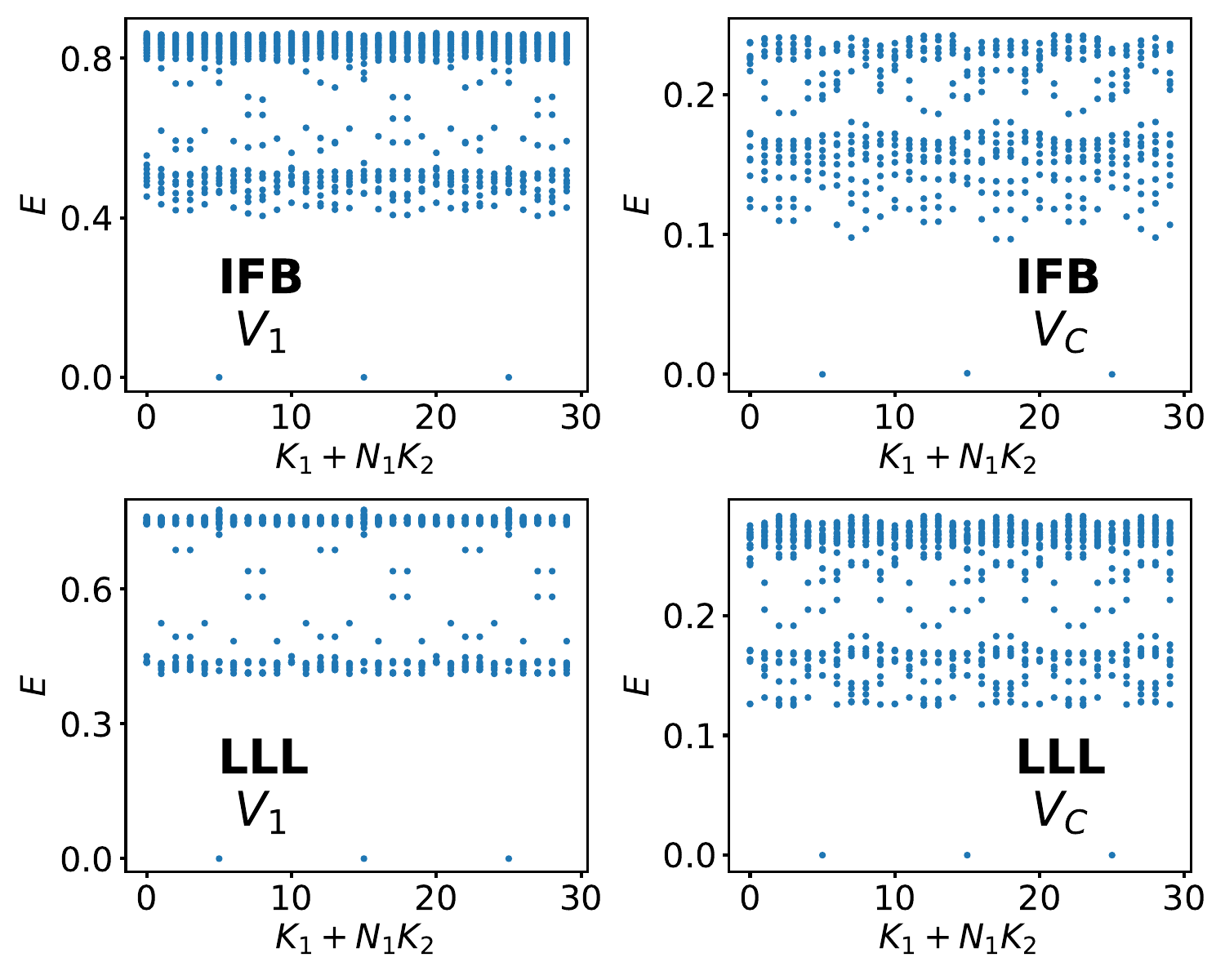}
\caption{\textbf{Spectra of different interactions at the $1/3$ filling in in moir\'e IFB and the LLL}.}
\label{fig_A1}
\end{figure}

In Fig.\ref{fig_A1}, the spectra of different interactions at the $1/3$ filling with $N_e = 10$ are provided, where the three-fold topological ground states can be observed. The ground state degeneracy is exact for the model Hamiltonian $V_1$ but approximate for Coulomb interactions. Such degeneracy is a signature of fractional Chern insulator (FCI) phases. Furthermore, with the same interaction, the magnetoroton modes have flatter dispersions in the lowest Landau level (LLL), and within the same band, the magnetoroton modes are flatter with the $V_1$ pseudopotential.

\begin{figure}[p]
\includegraphics[width=0.66\linewidth]{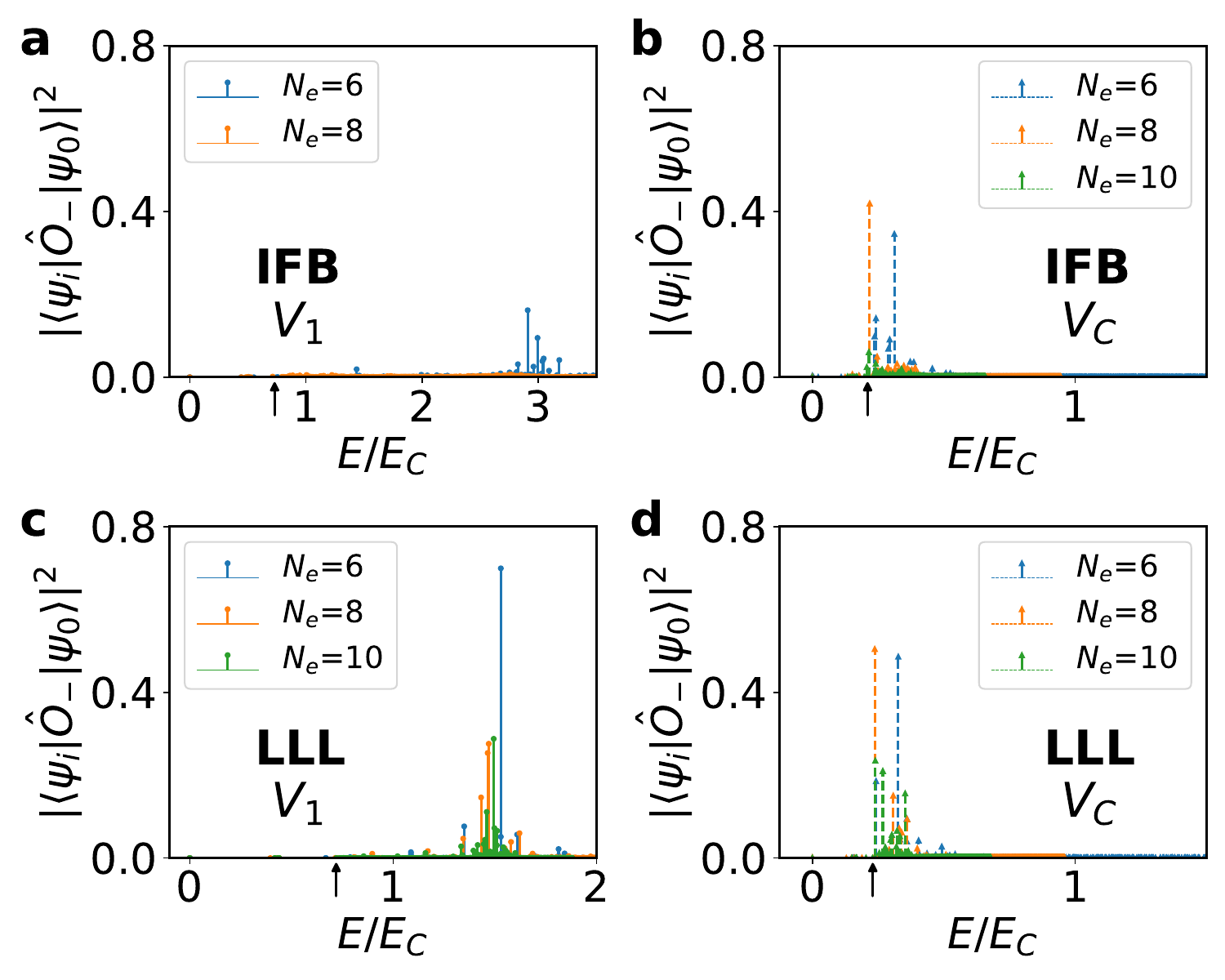}
\caption{\textbf{Spectral functions of the graviton operator with the chirality $\sigma = -$ at the filling $\nu = 1/3$ in moir\'e IFB and the LLL without fitting}. 
All the spectral peaks have been normalized. The system sizes are distinguished by different colors. Arrows denote the boundary positions of the excitation continuum. The energy scale is determined by the Coulomb energy $E_C= e^2/(4 \pi \epsilon l)$, where the characteristic length scale $l = \ell_B$ for LLL and $l = \sqrt{\sqrt{3} / 4 \pi} a_M$ for IFB. Here $a_M$ is the moir\'e lattice constant. }
\label{fig_A2}
\end{figure}

\begin{figure}[p]
\includegraphics[width=0.7\linewidth]{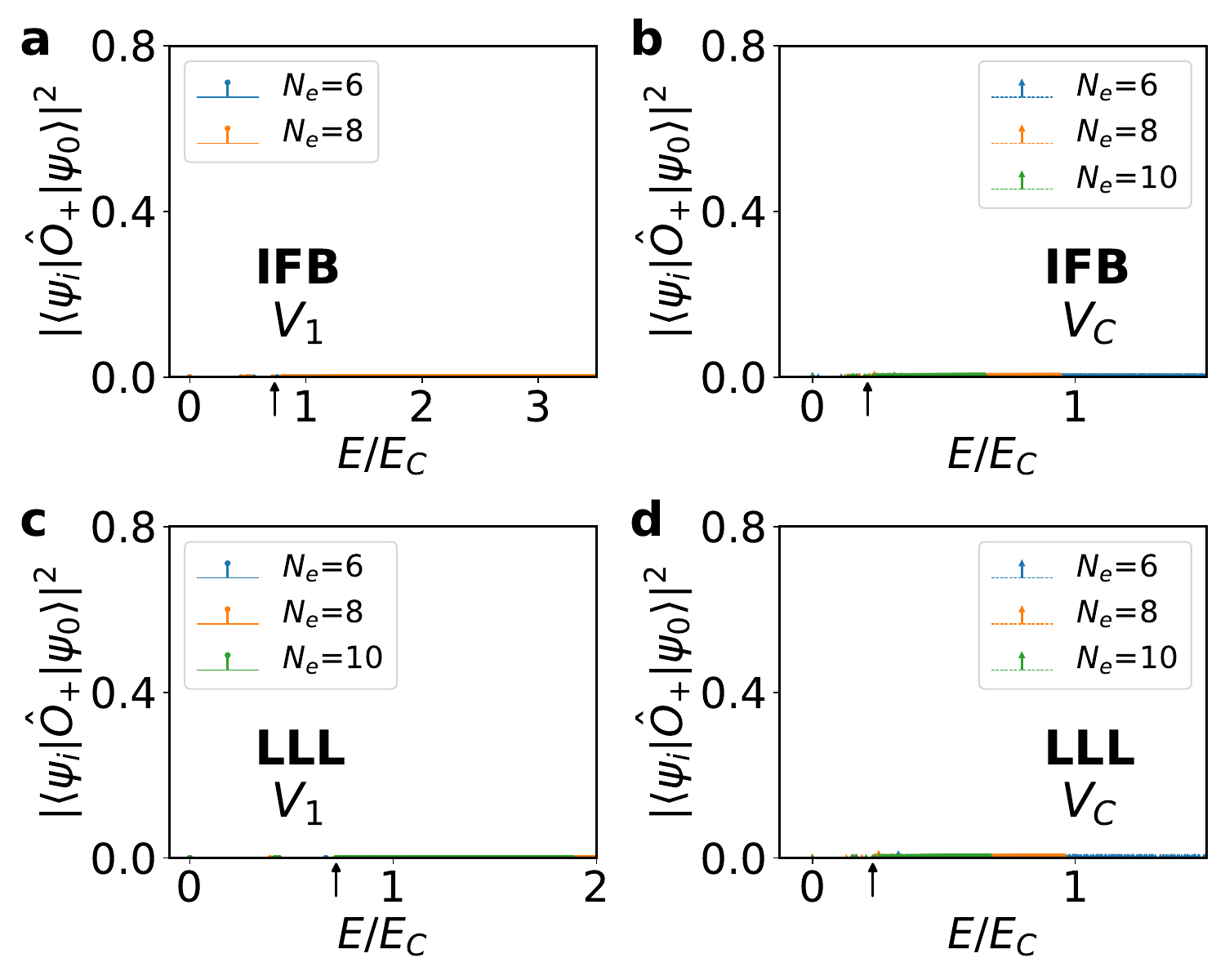}
\caption{\textbf{Spectral functions of the graviton operator with the chirality $\sigma = +$ at the filling $\nu = 1/3$ in moir\'e IFB and the LLL without fitting}. 
All the spectral peaks vanish, which have been normalized with the same normalization factor as the chirality $\sigma = -$. }
\label{fig_A2b}
\end{figure}

In Fig.\ref{fig_A2}, we show the spectral functions with chirality $\sigma = -$ without fitting, for the readers to clearly see how the GM peaks distribute in the spectrum. Furthermore, we also present the spectral functions of the positive chirality in Fig.~\ref{fig_A2b}, where one can see that the corresponding GMs are suppressed. 

\begin{figure}[p]
\includegraphics[width=0.7\linewidth]{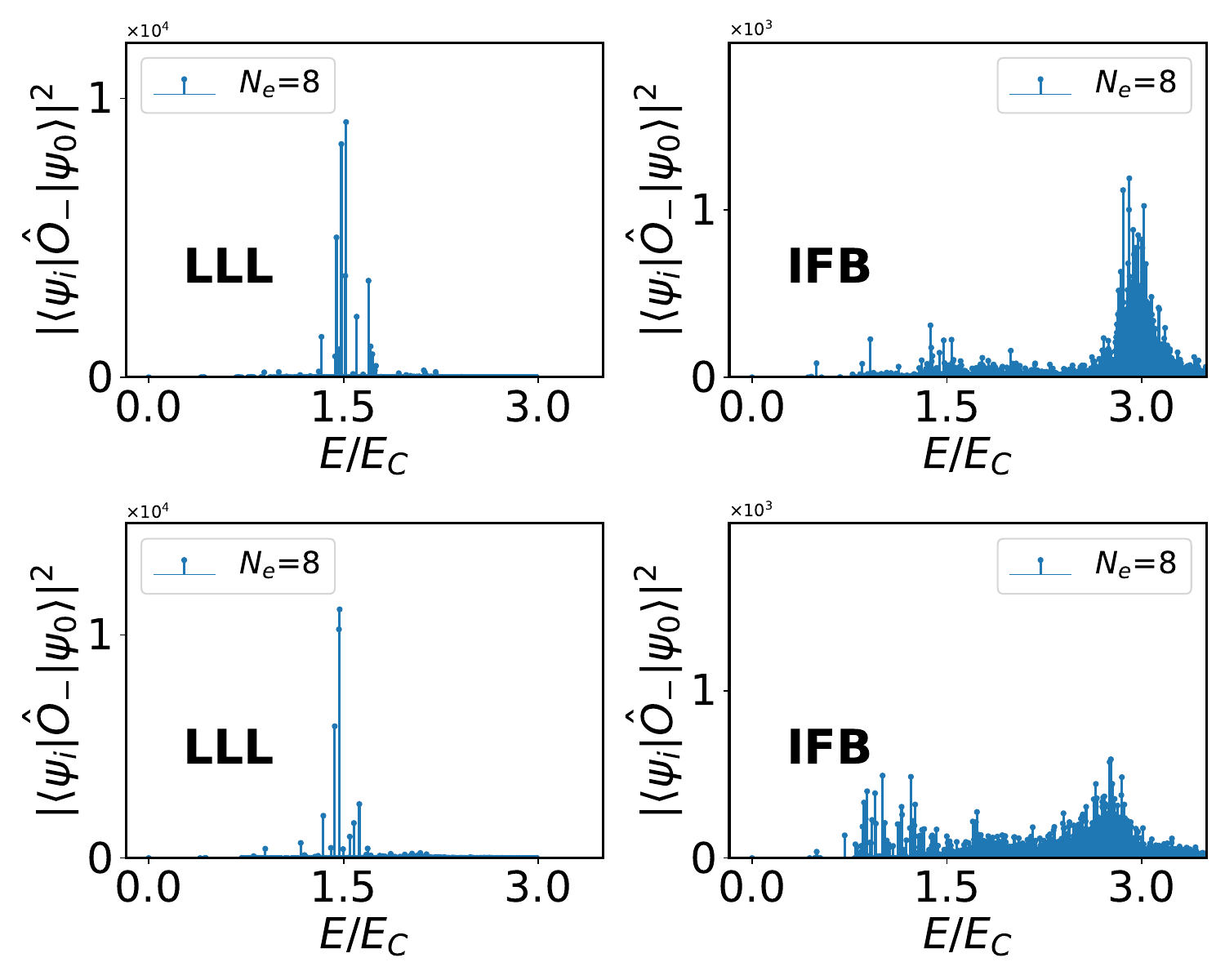}
\caption{\textbf{Spectral functions of $V_1$ without normalization}. Here in the top panels, the system size $(N_e,N_1,N_2)=(8,3,8)$, while in the bottom panels $(N_e,N_1,N_2)=(8,4,6)$. Here $N_1$ and $N_2$ are the number of unit cells along lattice vectors.}
\label{fig_A3}
\end{figure}

In Fig.\ref{fig_A3}, the spectral functions of $V_1$ for different sample shape without normalization are shown. Note that even if a seemingly high peak can be observed near the energy of GMs in the IFB cases, the exact values of the spectral peaks are extensive and have no physical meaning. Instead, the spectral functions only reveal how dispersed the GMs are.  Compared with the LLL at the same system size, one can see how they are dispersed into many more eigenstates around the GM energy in IFBs, implying a much shorter lifetime in the thermodynamic limit.

\begin{figure}[h]
\includegraphics[width=0.66\linewidth]{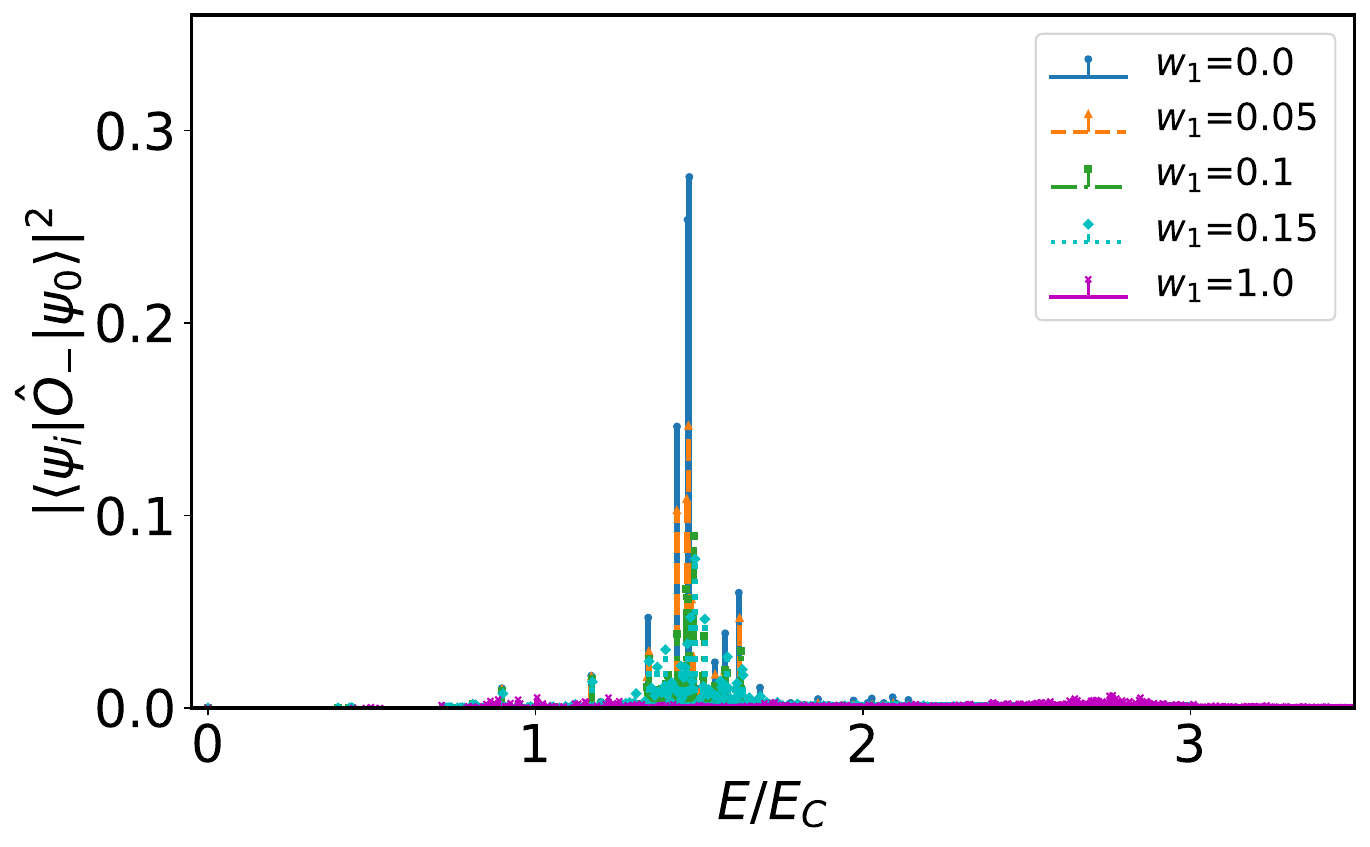}
\caption{\textbf{Spectral functions in IFBs with $N_e = 8$, $V_1$ interaction and different $w_1$}. }
\label{fig_A4}
\end{figure}

In Fig.\ref{fig_A4}, we show the effect of tuning $w_1$ on the spectral function of GMs, where one can observe how the GM peaks diminish monotonically with larger $w_1$, which gives a less uniform quantum geometry in the band.

In Table~\ref{tab:sm_A1}, we show the size of the overlap (in the form $|\langle \psi_1 | \psi_2 \rangle|^2$) between the ground states in the $(K_1, K_2)$ sector for $V_1$ and Coulomb interactions. The ground states in the system with $N_e = 6$ are within the same $\boldsymbol{k}$ sector in the Brillouin zone, so we instead calculated the total overlap between the sub-Hilbert spaces spanned by ground states. One can see that for numerically accessible system sizes, the overlap is quite high, which can be regarded as evidence of them belonging to the same topological phases. However, it is worth noticing that the overlap will vanish in the thermodynamic limit, because of the orthogonality catastrophe.

\begin{table}[h]
    \centering
    \small  
    \setlength{\tabcolsep}{6pt}  
    \renewcommand{\arraystretch}{1.4}  
    \begin{tabular}{ccccccc}
        \toprule
        $N_e$ & $N_1$ & $N_2$ & $K_1$ & $K_2$ & $w_1=0$ & $w_1=1$ \\
        \midrule
        6 & 3 & 6 & 0 & 3 & \multicolumn{2}{c}{Total = 1}  \\
        6 & 2 & 9 & 1 & 0 & 0.981 & 0.972\\
        8 & 3 & 8 & 0 & 4 & 0.966 & 0.948\\
        8 & 4 & 6 & 0 & 0  & 0.987 & 0.982\\
        10 & 5 & 6 & 0 & 1  & 0.976 & 0.962\\
        \bottomrule
    \end{tabular}
    \caption{\textbf{Overlap between the ground states of $V_1$ and Coulomb.} The overlap in the first case is for the subspace spanned by the three ground states.}
    \label{tab:sm_A1}
\end{table}

In Table~\ref{tab:sm_A2}, we show the overlap between the GMs in IFBs with $w_1=0$ (LLL) and $1$ (cTBG) with $V_1$ pseudopotential. Similar to the ground states, the overlap between the GMs in the LLL and cTBG IFBs is also very high for these system sizes.

\begin{table}[h]
    \centering
    \small  
    \setlength{\tabcolsep}{6pt}  
    \renewcommand{\arraystretch}{1.4}  
    \begin{tabular}{cccccc}
        \toprule
        $N_e$ & $N_1$ & $N_2$ & $K_1$ & $K_2$ & GM Overlap  \\
        \midrule
        8 & 4 & 6 & 0 & 0  & 0.8495 \\
        10 & 5 & 6 & 0 & 1  & 0.8474 \\
        \bottomrule
    \end{tabular}
    \caption{\textbf{Overlap between the GMs of IFBs with $w_1=0$ and $1$ with $V_1$.}}
    \label{tab:sm_A2}
\end{table}

\setcounter{table}{0}
\renewcommand{\thetable}{B.\arabic{table}}

\section{Chiral graviton operators} 
\label{sec:supp_B}

In this section, we derive the chiral graviton operators in the ideal flat band (IFB) and the continuum models. 

For Landau levels, to get the chiral graviton operator, we expand the Landau level form factor in terms of an effective mass anisotropic parameter $\varepsilon$ \cite{yang2013prb}:
\begin{equation}
\begin{aligned}
F_n(\boldsymbol{q}, a)^2
& =e^{-\frac{1}{2} \left(g \cdot q_x^2+ \frac{q_y^2}{g}\right) } \left|\mathcal{L}_n \left[ \frac{1}{2} \left(g \cdot q_x^2+\frac{q_y^2}{g}\right) \right]\right|^2, \quad g=1+\varepsilon
\end{aligned}
\end{equation}
where $\mathcal{L}(x)$ denotes the Laguerre polynomials, and we have set the magnetic length as unity. The system becomes isotropic with $\varepsilon=0$. We focus on the lowest Landau level with $n=0$, so the anisotropic Hamiltonian is written as:
\begin{equation}
\hat{H} =\frac{1}{2} \sum_{\boldsymbol{q}} V({\boldsymbol{q}}) e^{-\frac{1}{2} \left(g \cdot q_x^2+ \frac{q_y^2}{g}\right) } \hat{\bar{\rho}}_{\boldsymbol{q}} \hat{\bar{\rho}}_{-\boldsymbol{q}}
\end{equation}
where $\hat{\bar{\rho}}_{\boldsymbol{q}}$ denotes the guiding center density operators. For $1 / g$ we can expand it as $\frac{1}{g}=\frac{1}{1+\varepsilon} \approx 1-\varepsilon$. By ignoring higher-order terms in $\varepsilon$, one can substitute back into the exponent of the form factor:
\begin{equation}
-\frac{1}{2} \left[(1+\varepsilon) q_x^2+(1-\varepsilon) q_y^2\right] =-\frac{1}{2} \left[q_x^2+q_y^2+\epsilon\left(q_x^2-q_y^2\right)\right] = -\frac{1}{2} \left(q_x^2+q_y^2 \right) -\frac{\epsilon }{4} \left(\tilde{\boldsymbol{q}}^2+\tilde{\boldsymbol{q}}^{*2} \right),
\label{g_expansion}
\end{equation}
where $\tilde{\boldsymbol{q}} = q_x+ i q_y$. So the operator describing the leading-order coupling between an FQH state to the lattice distortion or metric fluctuation is given by:
\begin{equation}
\delta \hat{H}(\varepsilon)= \frac{\epsilon}{4} \sum_{\boldsymbol{q}}\left(q_y^2-q_x^2\right) V({\boldsymbol{q}}) e^{-\frac{1}{2} |\boldsymbol{q}|^2} \hat{\bar{\rho}}_{\boldsymbol{q}} \hat{\bar{\rho}}_{-\boldsymbol{q}}
\end{equation}
From this, we define the chiral graviton operators as:
\begin{equation}
\hat{O}^{\pm} = \sum_{\boldsymbol{q}} \left(q_x \pm i q_y\right)^2 V^{\prime}({\boldsymbol{q}})   \hat{\bar{\rho}}_{\boldsymbol{q}} \hat{\bar{\rho}}_{-\boldsymbol{q}}.
\label{LLL_opm}
\end{equation}
Note that as in Eq.~\ref{Opm} in the main text, here the Landau level form factor is absorbed into the interaction $V^{\prime}({\boldsymbol{q}}) = V({\boldsymbol{q}}) e^{-\frac{1}{2} |\boldsymbol{q}|^2}$. Furthermore, a non-chiral (linearly-polarized) graviton operator is defined as $\hat{O}^{(2)} = (\hat{O}^{+}+\hat{O}^{-})/2$. The factor $(q_x \pm i q_y)^2$ implies that such operator describes a quadrupole structure.

For the IFB case, since the explicit form factor can be derived, one can use exactly the same strategy to get the graviton operators. The second quantized form of the Hamiltonian is given by \cite{Wang2021}:
\begin{equation}
\hat{H}= \sum_{\boldsymbol{k}_1,\boldsymbol{k}_2,\boldsymbol{k}_3,\boldsymbol{k}_4} \left(\prod_{i=1}^4 \mathcal{N}_{k_i}\right) H^{\text{IFB}}_{\boldsymbol{k}_1 \boldsymbol{k}_2 ; \boldsymbol{k}_3 \boldsymbol{k}_{\mathbf{4}}} c_{\boldsymbol{k}_1}^{\dagger} c_{\boldsymbol{k}_2}^{\dagger} c_{\boldsymbol{k}_3} c_{\boldsymbol{k}_{\mathbf{4}}},
\end{equation}
with the matrix elements:
\begin{equation}
\begin{aligned}
& H^{\text{IFB}}_{\boldsymbol{k}_1 \boldsymbol{k}_2 ; \boldsymbol{k}_3 \boldsymbol{k}_4}= \sum_{\boldsymbol{b}} V(\boldsymbol{k}_1-\boldsymbol{k}_4-\boldsymbol{b}) \left(\sum_{\boldsymbol{b}_i} w_{\boldsymbol{b}_i} f_{\boldsymbol{b}-\boldsymbol{b}_i}^{\boldsymbol{k}_1, \boldsymbol{k}_4}\right)\left(\sum_{\boldsymbol{b}_j} w_{\boldsymbol{b}_j} f_{-\boldsymbol{b}+\delta \boldsymbol{b}-\boldsymbol{b}_j}^{\boldsymbol{k}_2, \boldsymbol{k}_3}\right),
\label{ifb_ham_ele}
\end{aligned}   
\end{equation}
where $\delta \boldsymbol{b}=\boldsymbol{k}_1+\boldsymbol{k}_2-\boldsymbol{k}_3-\boldsymbol{k}_4$, and the form factor is given by $f_{\boldsymbol{b}}^{\boldsymbol{k}, \boldsymbol{k}^{\prime}}=\eta_{\boldsymbol{b}} \cdot e^{\frac{i}{2}\left(\boldsymbol{k}+\boldsymbol{k}^{\prime}\right) \times \boldsymbol{b}} e^{\frac{i}{2} \boldsymbol{k} \times \boldsymbol{k}^{\prime}} e^{-\frac{1}{4}\left|\boldsymbol{k}-\boldsymbol{k}^{\prime}-\boldsymbol{b}\right|^2}$. We can see that $\boldsymbol{q}$ is substituted by $\boldsymbol{k}_1-\boldsymbol{k}_4-\boldsymbol{b}$ in the interaction, and the lowest Landau level form factor remains $e^{-\frac{1}{4}\left|\boldsymbol{q}\right|^2}$. (As in the LL case we only perturb the cyclotron coordinate metric within the Landau level form factor. One should distinguish this from $f_{\boldsymbol{b}}^{\boldsymbol{k}, \boldsymbol{k}^{\prime}}$.) Therefore, the expansion is exactly the same as in Eq.~\ref{g_expansion} and the final expression of the chiral graviton operator is only slightly more complicated because of the product rule. Let us define the chiral form factor as:
\begin{equation}
F_{\boldsymbol{b}}^{\pm, \boldsymbol{k}, \boldsymbol{k}^{\prime}} = \left[(k_x-k^\prime_x-b_x) \pm i (k_y-k^\prime_y-b_y)\right]^2 \cdot f_{\boldsymbol{b}}^{\boldsymbol{k}, \boldsymbol{k}^{\prime}}.
\end{equation}
The explicit expression of chiral graviton operators in IFBs is given by:
\begin{equation}
\begin{aligned}
\hat{O}_{\text{IFB}}^{\pm}= &\sum_{\boldsymbol{k}_1,\boldsymbol{k}_2,\boldsymbol{k}_3,\boldsymbol{k}_4} \left(\prod_{i=1}^4 \mathcal{N}_{k_i}\right) O^{\pm}_{\boldsymbol{k}_1 \boldsymbol{k}_2 ; \boldsymbol{k}_3 \boldsymbol{k}_{\mathbf{4}}} c_{\boldsymbol{k}_1}^{\dagger} c_{\boldsymbol{k}_2}^{\dagger} c_{\boldsymbol{k}_3} c_{\boldsymbol{k}_{\mathbf{4}}},\\
O^{\pm}_{\boldsymbol{k}_1 \boldsymbol{k}_2, \boldsymbol{k}_3 \boldsymbol{k}_{\mathbf{4}}} = & \sum_{\boldsymbol{b}} V(\boldsymbol{k}_1-\boldsymbol{k}_4-\boldsymbol{b}) 
\left[\left(\sum_{\boldsymbol{b}_i} w_{\boldsymbol{b}_i} F_{\boldsymbol{b}-\boldsymbol{b}_i}^{\pm, \boldsymbol{k}_1, \boldsymbol{k}_4}\right)\left(\sum_{\boldsymbol{b}_j} w_{\boldsymbol{b}_j} f_{-\boldsymbol{b}+\delta \boldsymbol{b}-\boldsymbol{b}_j}^{\boldsymbol{k}_2, \boldsymbol{k}_3}\right) \right.\\
& \qquad + \left. \left(\sum_{\boldsymbol{b}_i} w_{\boldsymbol{b}_i} f_{\boldsymbol{b}-\boldsymbol{b}_i}^{\boldsymbol{k}_1, \boldsymbol{k}_4}\right)\left(\sum_{\boldsymbol{b}_j} w_{\boldsymbol{b}_j} F_{-\boldsymbol{b}+\delta \boldsymbol{b}-\boldsymbol{b}_j}^{\pm, \boldsymbol{k}_2, \boldsymbol{k}_3}\right) \right]
\label{ifb_opm}
\end{aligned}
\end{equation}

To see whether these operators have well-defined chiralities, as shown in Table~\ref{tab:sm_B1}, we numerically confirmed the Laughlin graviton modes with positive chirality are completely annihilated when using the model Hamiltonian $V_1$ in the lowest Landau level and the cTBG IFB. For the Coulomb interaction, even if the chirality is not as exact as the model Hamiltonian, the negative chirality still overwhelms the positive one. Furthermore, while Eq.~\ref{ifb_opm} is the most general expression of the graviton operators, numerically we found that Eq.~\ref{ifb_opm} can be simplified into the same form as Eq.~\ref{LLL_opm}, i.e., the chiral graviton operators are given by multiplying $\tilde{\boldsymbol{q}}$ or $\tilde{\boldsymbol{q}}^*$ to the interaction $V(\boldsymbol{q})$ in the Hamiltonian without modifying the form factors.

\begin{table}[h]
    \centering
    \small  
    \setlength{\tabcolsep}{5pt}  
    \renewcommand{\arraystretch}{1.2}  
    \begin{tabular}{ccccccccc}
        \toprule
        \multicolumn{9}{c}{Ideal flat band models ($w_1 = 0$: LLL; $w_1 = 1$: cTBG)} \\
        \midrule
        $N_e$ & $N_1$ & $N_2$ & Interaction & $w_1$ & $\hat{V}_1$ & $\hat{O}^{+}$ & $\hat{O}^{-}$ & $\hat{O}^{(2)}$  \\
        \midrule
        \multirow{4}{*}{8} & \multirow{4}{*}{3} & \multirow{4}{*}{8} & \multirow{2}{*}{$V_1$} & 0 & $10^{-10}$ & $10^{-8}$ & 201.0 & 100.5  \\
        & & & & 1 & $10^{-4}$ & 0.008 & 369.1 & 184.5 \\
        \cmidrule{4-9}
        & & & \multirow{2}{*}{Coulomb} & 0 & - & 16.1 & 192.8 & 96.4  \\
        & & & & 1 & - & 34.4 & 352.3 & 123.5 \\
        \midrule
        \multirow{4}{*}{8} & \multirow{4}{*}{4} & \multirow{4}{*}{6} & \multirow{2}{*}{$V_1$}  & 0 & $10^{-10}$ & $10^{-8}$ & 201.0 & 100.5 \\
        & & & & 1 & $10^{-4}$ & 0.008 & 301.2 & 150.6 \\
        \cmidrule{4-9}
        & & & \multirow{2}{*}{Coulomb} & 0 & - & 8.2 & 197.1 & 98.4  \\
        & & & & 1 & - & 14.4 & 294.7 & 147.2 \\
        \bottomrule
    \end{tabular}
    \caption{\textbf{Normalization factors of acting the operators on the ground state.} The normalization factor of acting $\hat{V}_1$ on the ground state estimates the numerical errors in the calculations, since ideally it should vanish.}
    \label{tab:sm_B1}
\end{table}

For the continuum models, since the form factors have to be numerically calculated from the Bloch states, we instead take the quadrupole operator constructed from the $L=2$ (d-wave) representation of the lattice symmetry group as the graviton operators. These operators will automatically fulfill the lattice symmetry and the boundary conditions.

Here we use the dihedral group $D_3$ as an example, which captures the twisted MoTe$_2$ lattice symmetry. The group $D_3$ has six elements:$\{\, e,\, C_3(z),\, C_3^2(z),\, C_2,\, C_2^2,\, C_2^3 \}$, where $e$ is the identity, and $C_3(z)$ and $C_3^2(z)$ are rotations by $\pm 2\pi/3$ along $z$ axis at the center; The other ones are rotations by $\pi$ about the lines through vertices and midpoints of a triangle. The character table is provided in Table.~\ref{tab:D3}.

\begin{table}[h]
    \centering
    \small  
    \setlength{\tabcolsep}{8pt}  
    \renewcommand{\arraystretch}{1.4}  
    \begin{tabular}{c|ccc|c|c}
        \toprule
        $\mathbf{D_3}$ & $e$ & $2C_3(z)$ & $3C_2$ & Linear functions & Quadratic functions \\
        \midrule
        $A_1$ & +1 & +1 & +1 & -- & $x^2 + y^2, z^2$ \\
        $A_2$ & +1 & +1 & -1 & $z$ & -- \\
        $E$   & +2 & -1 & 0 & $(x, y)$ & $(x^2 - y^2, xy), (xz, yz)$ \\
        \bottomrule
    \end{tabular}
    \caption{\textbf{Character table of $D_3$ with linear and quadratic function representations.}}
    \label{tab:D3}
\end{table}

For a 2D triangular lattice, we can choose three fundamental displacement vectors $\boldsymbol{a}_1,\boldsymbol{a}_2,\boldsymbol{a}_3$ such that:
\begin{equation}
\boldsymbol{a}_1 + \boldsymbol{a}_2 + \boldsymbol{a}_3 = 0,
\end{equation}
with each pair $|\boldsymbol{a}_i - \boldsymbol{a}_j|$ being the same length. Without loss of generality, one can use:
\begin{equation}
\boldsymbol{a}_1 = \left(1,\,0\right), \quad
\boldsymbol{a}_2 = \left(-\tfrac12,\, \tfrac{\sqrt{3}}{2}\right), \quad
\boldsymbol{a}_3 = \left(-\tfrac12,\, -\tfrac{\sqrt{3}}{2}\right).
\end{equation}
So $\theta_1 = 0$, $\theta_2 = 2 \pi/3$, $\theta_3 = 4 \pi/3$, i.e., they are $2\pi/3$ apart. In a honeycomb lattice, each primitive cell may contain two sites, but the same $D_3$ point-group geometry applies to the underlying Bravais vectors. With the fundamental displacement vectors, we define three real functions that depend on a wavevector $\boldsymbol{q}$:
\begin{equation}
\psi_1(\boldsymbol{q}) = \cos(\boldsymbol{q}\cdot \boldsymbol{a}_1), \quad
\psi_2(\boldsymbol{q}) = \cos(\boldsymbol{q}\cdot \boldsymbol{a}_2), \quad
\psi_3(\boldsymbol{q}) = \cos(\boldsymbol{q}\cdot \boldsymbol{a}_3).
\end{equation}
They form a basis for a real representation (fulfilling periodic boundary conditions) of $D_3$, which has three irreducible representations $A_1, A_2,E,$ as shown in Table.~\ref{tab:D3}. Here $A_1$ and $A_2$ are one-dimensional, and $E$ is a two-dimensional representation. We can find the explicit $E$-representation by seeking linear combinations of $(\psi_1,\psi_2,\psi_3)$ that remain orthogonal to the $A_1$ representation, and match the angular dependence of $(x^2 - y^2, xy) \sim (\cos(2 \theta_i), \sin (2 \theta_i))$. The representations of these operators turn out to be:
\begin{equation}
\begin{aligned}
\mathcal{O}_{x^2 - y^2}(\boldsymbol{q})
&= 2 \cdot \cos(\boldsymbol{q}\cdot \boldsymbol{a}_1)
      -\cos(\boldsymbol{q}\cdot \boldsymbol{a}_2)
      -\cos(\boldsymbol{q}\cdot \boldsymbol{a}_3),
\\[6pt]
\mathcal{O}_{xy}(\boldsymbol{q})
&= \sqrt{3} \cdot [\cos(\boldsymbol{q}\cdot \boldsymbol{a}_3) -\cos(\boldsymbol{q}\cdot \boldsymbol{a}_2)].
\end{aligned}
\end{equation}
The graviton operators are then defined as:
\begin{equation}
\hat{O}_{t\text{MoTe}_2}^{\pm} = \sum_{\boldsymbol{q}} \left(\mathcal{O}_{x^2 - y^2} \pm i \mathcal{O}_{xy}\right) V({\boldsymbol{q}})   \hat{\tilde{\rho}}_{\boldsymbol{q}} \hat{\tilde{\rho}}_{-\boldsymbol{q}}.
\label{mote2_opm}
\end{equation}
Here $\hat{\tilde{\rho}}_{\boldsymbol{q}} $ denotes the projected density operator on the Chern band. All the numerical results of the continuum models in this work are based on Eq.~\ref{mote2_opm}.

\section{Generalized pseudopotentials} 
\label{sec:supp_C}
In this section, we introduce the definition of generalized pseudopotentials, which were first considered in Ref.~\cite{yang2017generalized}. By defining the center-of-mass and relative guiding center coordinates $\bar{R}^{a}_{i j}= R_i^a+R_j^a, R^{a}_{i j}=R_i^a-R_j^a, a \in \{ x, y\}$, one can write down the two-body eigenstates for a pair of electrons $|M, m\rangle$, where $M$ is the index for $\hat{\bar{R}}_{i j}$, and $m$ is the index for $\hat{R}_{i j}$ (For bosons/fermions, $m$ can only be even/odd). For interactions $V(r)$ that only depend on the relative coordinates $r_i - r_j$ (such as the Coulomb interaction $V_C \sim 1/|r_i - r_j|$), $M$ is conserved, so we can get the Fourier transform $V_q$ and write down the Hamiltonian:
\begin{equation}
\hat{\mathcal{H}}_{2 \mathrm{bdy}}=\sum_{i \neq j} 
 \sum_{M, m, m^{\prime}}  \left(\int \frac{d^2 q}{(2 \pi)^2} V_q \langle M, m^{\prime}| e^{i q_a \left(\hat{R}_i^a-\hat{R}_j^a\right)}\left|M, m\right\rangle \right) \left|M, m^{\prime}\right\rangle \langle M, m|,
\end{equation}
where we omit the particle indices in $\left|M_{ij}, m_{ij}\right\rangle$ and absorb the Landau level form factor into $V_q$. The matrix element can be calculated by defining the ladder operator $\hat{a} = \hat{R}_{ij}^x + i \hat{R}_{ij}^{y}$ and the complex variable $\tilde{\boldsymbol{q}} \equiv(1 / \sqrt{2})\left(q_x+i q_y\right)$, using the Baker–Campbell–Hausdorff formula:
\begin{equation}
\exp \left[i \sqrt{2} \tilde{\boldsymbol{q}} \hat{a}^{\dagger}+ i \sqrt{2} \tilde{\boldsymbol{q}}^{*} \hat{a}\right]= e^{-\frac{1}{2}| \boldsymbol{q}|^2} \exp \left[i \sqrt{2} \tilde{\boldsymbol{q}} \hat{a}^{\dagger}\right] \exp \left[i \sqrt{2} \tilde{\boldsymbol{q}}^{*} \hat{a}\right]=e^{-\frac{1}{2}| \boldsymbol{q}|^2} \sum_{r=0}^{\infty} \sum_{s=0}^{\infty} \frac{1}{r!s!}\left(i \sqrt{2} \tilde{\boldsymbol{q}} \hat{a}^{\dagger}\right)^r\left(i \sqrt{2} \tilde{\boldsymbol{q}}^{*} \hat{a}\right)^s,
\end{equation}
so we have
\begin{equation}
\begin{aligned}
\langle m^{\prime}| \exp \left[i \sqrt{2} \tilde{\boldsymbol{q}} \hat{a}^{\dagger}\right] \exp \left[i \sqrt{2} \tilde{\boldsymbol{q}}^{*} \hat{a}\right]\left|m\right\rangle= & \sum_{r=0}^{\infty} \sum_{s=0}^{\infty} \frac{1}{r!s!}\left(i \sqrt{2} \tilde{\boldsymbol{q}}\right)^r\left(i \sqrt{2} \tilde{\boldsymbol{q}}^{*}\right)^s\langle m^{\prime}|\left(\hat{a}^{\dagger}\right)^r (\hat{a})^s\left|m\right\rangle\\
=& \sum_{r=0}^{\infty} \sum_{s=0}^{\infty} \frac{1}{r!s!}\left(i \sqrt{2} \tilde{\boldsymbol{q}}\right)^r\left(i \sqrt{2} \tilde{\boldsymbol{q}}^{*}\right)^s \delta_{m^{\prime}, m-s+r} \sqrt{\frac{m!}{\left(m-s\right)!}} \sqrt{\frac{\left(m-s+r\right)!}{\left(m-s\right)!}}.
\end{aligned}
\end{equation}
Finally, we note that such a series defines the generalized Laguerre polynomials $\mathcal{L}_m^n$:
\begin{equation}
\langle M, m^{\prime}| e^{i q_a\left(R_1^a-R_2^a\right)}\left|M, m\right\rangle=\sqrt{\frac{m!}{m^{\prime}!}}\left(i \sqrt{2} \tilde{\boldsymbol{q}}\right)^{\Delta m} e^{-\frac{1}{2}| \boldsymbol{q}|^2} \mathcal{L}_m^{\Delta m}\left(|\boldsymbol{q}|^2\right),
\label{gen_pp}
\end{equation}
where $|\boldsymbol{q}|^2 /2 = \tilde{\boldsymbol{q}}\tilde{\boldsymbol{q}}^{*}$ and we define $\ell_B \equiv 1$. $\Delta m \equiv m^{\prime}-m$ ($\Delta m$ is assumed to be positive above without loss of generality). We define Eq.~\ref{gen_pp} and its conjugate as the generalized pseudopotentials $V_{m, \Delta m} ^{+}$ and $V_{m, \Delta m} ^{-}$. So for the interactions holomorphic in $\boldsymbol{q}$ as in the perturbation terms in $\hat{H}_s$ in the main text, the expansion can only involve terms with $V_{m, \Delta m} ^{+}$. Again, only the components with odd $m$ are relevant for fermions. Thus for the Laughlin state at the filling $1/n$, the relative angular momentum between any two electrons is greater than $n-2$, so the holomorphic generalized pseudopotential $\sum_m c_m |m\rangle \langle n^\prime|$ with $n^\prime < n$ acting on all pairs of electrons (as in Eq.~15 in the main text, where we focus on the $n=3$ case) will naturally annihilate this Laughlin state. 

Note that if the interaction is isotropic, $m$ is also conserved. So we can expand the interaction with the Haldane pseudopotential $V_m$:
\begin{equation}
\langle M, m| e^{i q_a\left(R_1^a-R_2^a\right)}\left|M, m^{\prime}\right\rangle=\delta_{m, m^{\prime}} e^{-\frac{1}{2} |\boldsymbol{q}|^2} \mathcal{L}_m\left(|\boldsymbol{q}|^2\right) \equiv \delta_{m, m^{\prime}} V_m.
\end{equation}

\section{Hermiticity of the Hamiltonian} 
\label{sec:supp_D}

In this section, we provide a simple criterion for maintaining the Hermiticity in the Hamiltonian of ideal flat bands. By taking the conjugate of the form factor
\begin{equation}
f_{\boldsymbol{b}}^{\boldsymbol{k} \boldsymbol{k}^{\prime}} =\eta_{\boldsymbol{b}} e^{\frac{i}{2}\left(\boldsymbol{k}+\boldsymbol{k}^{\prime}\right) \times \boldsymbol{b}} e^{\frac{i}{2} \boldsymbol{k} \times \boldsymbol{k}^{\prime}} e^{-\frac{1}{4}\left|\boldsymbol{k}-\boldsymbol{k}^{\prime}-\boldsymbol{b}\right|^2},
\end{equation}
we can get:
\begin{equation}
(f^{\boldsymbol{k}_1\boldsymbol{k}_4}_{\boldsymbol{b}-\boldsymbol{b}_i})^* = f^{\boldsymbol{k}_4\boldsymbol{k}_1}_{-\boldsymbol{b}+\boldsymbol{b}_i}.
\end{equation}
Thus when we introduce another summation over $\boldsymbol{b}_i \in sRLV$, as long as for any $ \boldsymbol{b}_i$ in the summation, $-\boldsymbol{b}_i$ is also summed over, the Hamiltonian will stay Hermitian (The Hamiltonian matrix element of ideal flat bands can be found in Eq.~\ref{ifb_ham_ele}). So the simplest Hermitian toy model is to keep only one form factor in the linear term of $w_1$, and sum over a pair of inverse vectors in $sRLV$. The matrix element is given by:
\begin{equation}
h_{\boldsymbol{k}_1 \boldsymbol{k}_2 ; \boldsymbol{k}_3 \boldsymbol{k}_4}=\sum_{\boldsymbol{b}} v_{\boldsymbol{k}_1-\boldsymbol{k}_4-\boldsymbol{b}} \cdot\left[\left(f_{\boldsymbol{b}}^{\boldsymbol{k}_1, \boldsymbol{k}_4}+ w_1 \sum_{\boldsymbol{b}_i= \pm \boldsymbol{B}_1} f_{\boldsymbol{b}-\boldsymbol{b}_i}^{\boldsymbol{k}_1, \boldsymbol{k}_4}\right) \cdot f_{-\boldsymbol{b}+\delta \boldsymbol{b}}^{\boldsymbol{k}_2, \boldsymbol{k}_3}\right].
\end{equation}

\section{Projection-invariant holomorphicity of interactions}
\label{sec:supp_E}

In this section, we show that any interaction holomorphic in $\tilde{\boldsymbol{q}}$ remains holomorphic upon projection to either the lowest Landau level or a lattice Chern band, and that the converse is also true. Here the holomorphicity means that the total interaction is given by some bare interaction (such as a $V_1$ pseudopotential) multiplied by a holomorphic function of $\tilde{\boldsymbol{q}}$ (with the form factors of the relevant band considered separately to the interaction). In particular, for a generic interaction Hamiltonian in the full Hilbert space (see Eq.~\ref{gen_ham_E1}), projecting into any band simply multiplies single‐particle form factors, just like a density–density interaction. While the specific form of the interaction in Landau‐level projections may differ by a gauge-dependent phase, its holomorphicity is nevertheless preserved.

As we explain later in this section, this preservation of holomorphicity allows the generalized pseudopotential expansion that we describe in Section V to transfer to an effective real space interaction, allowing us to understand the properties of the interacting phase. If the interaction projected into a Chern band is holomorphic and only contains $V_{i,m}^{+}$ components \textit{after factoring out the Landau level form factors}, the Laughlin states at the filling $\nu=1/n$ remain zero‐energy states with emergent guiding‐center rotational symmetry as long as $i<n$. If this interaction does not generate negative‐energy states, thereby ensuring that the system hosts an FCI Laughlin phase, the graviton modes in these systems acquire substantially shorter lifetimes when they are within the excitation continuum compared to those in Landau levels. For more general Chern bands and more realistic interactions, where the projected interaction may contain other components, comparing the relative sizes of the components may allow us to determine how close the resulting system is to this idealized case.

To prove that the holomorphicity of interactions is invariant under projections, we consider a generic Hamiltonian:
\begin{equation}
\hat{H}_{2B}=\int d^2 \boldsymbol{q} \iint d^2 \boldsymbol{\kappa}_1 d^2 \boldsymbol{\kappa}_2 \tilde{V}_{\boldsymbol{q}, \boldsymbol{\kappa}_1, \boldsymbol{\kappa}_2}  \hat{c}_{\boldsymbol{\kappa}_1+\boldsymbol{q}}^{\dagger} \hat{c}_{\boldsymbol{\kappa}_2-\boldsymbol{q}}^{\dagger} \hat{c}_{\boldsymbol{\kappa}_1} \hat{c}_{\boldsymbol{\kappa}_2}.
\label{gen_ham_E1}
\end{equation}
Here all the integrals are carried out on the whole plane $\mathbb{R}^2$. $\hat{c}^{\dagger}_{\boldsymbol{\kappa}}/\hat{c}_{\boldsymbol{\kappa}}$ are the plane wave creation/annihilation operators, and the interaction is periodic across Brillouin zones (BZs) in the reciprocal space:
\begin{equation}
\tilde{V}_{\boldsymbol{q}, \boldsymbol{\kappa}_1+\boldsymbol{G}_1, \boldsymbol{\kappa}_2+\boldsymbol{G}_2}=\tilde{V}_{\boldsymbol{q}, \boldsymbol{\kappa}_1, \boldsymbol{\kappa}_2}.
\label{int_period}
\end{equation}
$\boldsymbol{G}_i$ here and $\boldsymbol{G}$, $\boldsymbol{g}$, $\boldsymbol{g}^\prime$ below are all the reciprocal lattice vectors. One can see that generically $\hat{H}_{2B}$ does not describe a density-density interaction.

For the lattice Chern bands, the eigenstates are the normal Bloch states $| \psi_{\boldsymbol{k},\text{Bloch}} \rangle$ obeying:
\begin{equation}
\left\langle\boldsymbol{\kappa} | \psi_{\boldsymbol{k}, \text{Bloch}}\right\rangle=(2 \pi)^2 \sum_{\boldsymbol{G}} c_{\boldsymbol{k}}(\boldsymbol{G}) \delta^{(2)}[\boldsymbol{\kappa}-(\boldsymbol{k}+\boldsymbol{G})].
\end{equation}
We can show that projecting to the band simply results in the multiplication by a form factor, just as it would for a standard density-density interaction. Because we exclude the form factor from the definition of the interaction, this does not change the holomorphicity in the transformed interaction $\tilde{V}_{\boldsymbol{q}, \boldsymbol{k}_1, \boldsymbol{k}_2}$. That is, if the interaction is holomorphic before projection, it is also holomorphic afterwards and vice-versa.

For the Landau levels, we use the magnetic Bloch states $|\phi_{k_x,k_y} \rangle$ which are the eigenstates of a set of commutative magnetic translation operators. The overlap between $|\phi_{k_x,k_y} \rangle$ and a plane wave $\boldsymbol{\kappa}$ can be written as
\begin{equation}
\begin{aligned}
\langle \boldsymbol{\kappa} | \phi_{k_x, k_y} \rangle
= & 2 \sqrt{2} \pi^{\frac{5}{4}}  \ell  e^{-\frac{1}{2}\left(\kappa_y \ell\right)^2} e^{-i \kappa_y \ell^2 k_x} e^{i (\kappa_x-k_x) \cdot \ell^2 \cdot (k_y -\kappa_y)} \sum_{g} \delta(g-(\kappa_x-k_x)).
\end{aligned}
\end{equation}
where we have chosen a Landau gauge $\boldsymbol{A}=B(-y, 0,0)$ without loss of generality, and the magnetic length is defined as $\ell=\sqrt{\frac{\hbar c}{e B}}$. Then the projected Hamiltonian in second-quantized form is given by:
\begin{equation}
\begin{aligned}
\hat{H}^{\prime}_{2B}= & \sum_{\{\boldsymbol{k}_i\}}
\left[\int d^2 \boldsymbol{q} \
\sum_{\boldsymbol{g}} e^{+i q_y \ell^2 (k_{1,x} - k_{2,x})}  \cdot e^{- i \left( q_x - k_{1, x} + k_{3, x}\right) \cdot \ell^2 \cdot (k_{1, y}  - q_y)}  
\delta^{(2)}(\boldsymbol{q}  - \boldsymbol{k}_{1}+ \boldsymbol{k}_{3} - \boldsymbol{g}) \right.\\
& \qquad  \cdot \sum_{\boldsymbol{g}^\prime} e^{- i \left( -q_x - k_{2, x} + k_{4, x}\right) \cdot \ell^2 \cdot (k_{2, y}  + q_y)}  e^{-\left(q_y  \ell\right)^2} 
\delta^{(2)}(-\boldsymbol{q}  - \boldsymbol{k}_{2}+ \boldsymbol{k}_{4} - \boldsymbol{g}^\prime) \\
& \qquad \cdot  \left.
\iint_{-\infty}^{\infty} d \kappa_{1,y} d \kappa_{2,y} 
\tilde{V}_{\boldsymbol{q}, (k_{3, x}, \kappa_{1,y}), (k_{4, x}, \kappa_{2,y})} 
e^{- (\kappa_{1,y}^2 +\kappa_{2,y}^2)\ell^2} 
e^{ i \tilde{\boldsymbol{q}} \cdot (\kappa_{1,y}-\kappa_{2,y} )\cdot \ell^2 }\right]
\hat{d}_{\boldsymbol{k}_1}^{\dagger} \hat{d}_{\boldsymbol{k}_2}^{\dagger}
\hat{d}_{\boldsymbol{k}_3}
\hat{d}_{\boldsymbol{k}_4},
\end{aligned}
\end{equation}
where $\hat{d}^{\dagger}_{\boldsymbol{\kappa}}/\hat{d}_{\boldsymbol{\kappa}}$ are the creation/annihilation operators for the magnetic Bloch states. The expression on the first two lines is a product of Landau level form factors, which we exclude from the definition of the projected interaction. The projected interaction is instead given by the integral in the third line. Since the integrand is holomorphic in $\tilde{\boldsymbol{q}}$, the projected interaction has the same holomorphicity as the unprojected interaction. That is, if the unprojected interaction is given by the bare interaction, such as $V_1(\boldsymbol{q}) \propto (1- |\boldsymbol{q}|^2)$, multiplied by a holomorphic function of $q$, then the projected interaction is the same bare interaction multiplied by some (generally different) holomorphic function. Similarly, if the projected interaction has this holomorphic form, the unprojected interaction has the same form (although, again, the exact form of the interaction may become different).

These results tell us that, given a projected Hamiltonian that is holomorphic (such as $H_s$ from Section V, which is projected to the LLL), we can write down an interaction in the plane wave basis that shares this holomorphicity. This unprojected interaction can then be expanded in terms of the generalized pseudopotentials that we described in Section \ref{sec:supp_C}. The Hamiltonian in Eq.~\ref{gen_ham_E1} can be written in the spatial basis as
\begin{align}
	\hat{H}_{2B}= \int d^2r_1 d^2r_2 d^2r_3d^2r_4 V_{\boldsymbol{r}_1- \boldsymbol{r}_2, \boldsymbol{r}_3, \boldsymbol{r}_4} c^{\dagger}_{\boldsymbol{r}_1+ \boldsymbol{r}_3} c^{\dagger}_{\boldsymbol{r}_2 + \boldsymbol{r}_4} c_{\boldsymbol{r}_1} c_{\boldsymbol{r}_2},
\end{align}
 where
\begin{align}
	V_{\boldsymbol{r}_1- \boldsymbol{r}_2, \boldsymbol{r}_3, \boldsymbol{r}_4} = \int d^2 q \int d^2 \kappa_1 d^2 \kappa_2 e^{i \boldsymbol{q} \cdot (\boldsymbol{r}_1 - \boldsymbol{r}_2) } e^{i \boldsymbol{\kappa}_1 \cdot \boldsymbol{r}_3}  e^{i \boldsymbol{\kappa}_2 \cdot \boldsymbol{r}_4} V_{\boldsymbol{q}, \boldsymbol{\kappa}_1, \boldsymbol{\kappa}_2},
\end{align}
We note that the annihilation operators only depend on the co-ordinates $\boldsymbol{r}_1$ and $\boldsymbol{r}_2$. As a result, the dependence of $V_{\boldsymbol{r}_1- \boldsymbol{r}_2, \boldsymbol{r}_3, \boldsymbol{r}_4}$ on $\boldsymbol{r}_1-\boldsymbol{r}_2$ will determine if a state is annihilated by the interaction, and hence will determine the existence of zero-energy states. This $\boldsymbol{r}_1$ and $\boldsymbol{r}_2$ dependence is entirely encoded within the $\boldsymbol{q}$ dependence. We can then expand $V_{\boldsymbol{q}, \boldsymbol{\kappa}_1, \boldsymbol{\kappa}_2}$ in terms of the generalized pseudopotentials from Eq.~\ref{gen_pp} (excluding the Gaussian factor in Eq.~\ref{gen_pp}, which is from the LLL form factor):
$$V_{\boldsymbol{q}, \boldsymbol{\kappa}_1, \boldsymbol{\kappa}_2}= \sum_{m,\Delta m } C_{\boldsymbol{\kappa}_1, \boldsymbol{\kappa}_2, m, \Delta m} V^+_{m, \Delta m}(\boldsymbol{q}),$$
where only $+$ pseudopotentials appear due to the holomorphicity. In the case of $H_s$, where we take $V(\boldsymbol{q})=V_1$, this expansion will only include contributions from $m=0$ and $m=1$. With this expansion, we then have
\begin{align}
	V_{\boldsymbol{r}_1- \boldsymbol{r}_2, \boldsymbol{r}_3, \boldsymbol{r}_4} &=  \sum_{m,\Delta m } \int d^2 \kappa_1 d^2 \kappa_2 e^{i \boldsymbol{\kappa}_1 \cdot \boldsymbol{r}_3}  e^{i \boldsymbol{\kappa}_2 \cdot \boldsymbol{r}_4} C_{\boldsymbol{\kappa}_1, \boldsymbol{\kappa}_2, m, \Delta m} \int d^2 q V^+_{m, \Delta m}(\boldsymbol{q}) e^{i \boldsymbol{q} \cdot (\boldsymbol{r}_1 - \boldsymbol{r}_2) } \notag \\
	&= \sum_{m,\Delta m } \int d^2 \kappa_1 d^2 \kappa_2 e^{i \boldsymbol{\kappa}_1 \cdot \boldsymbol{r}_3}  e^{i \boldsymbol{\kappa}_2 \cdot \boldsymbol{r}_4} C_{\boldsymbol{\kappa}_1, \boldsymbol{\kappa}_2, m, \Delta m} \tilde{V}^+_{m, \Delta m}(\boldsymbol{r}_1 - \boldsymbol{r}_2),
\end{align}
where $\tilde{V}^+_{m, \Delta m}(\boldsymbol{r}_1 - \boldsymbol{r}_2)$ is the spatial representation of the pseudopotential. Given that the only contributions are from the $m=0$ and $m=1$ pseudopotentials, this interaction annihilates the Laughlin-like states, as we described in Section \ref{sec:supp_C}. This guarantees that the Laughlin-like states are also exact zero-energy states of the projected interaction $H_s$. We note that, while the $\boldsymbol{r}_1-\boldsymbol{r}_2$ dependence guarantees the existence of these zero-energy states, the higher energy eigenstates will also depend on the $\boldsymbol{r}_3$ and $\boldsymbol{r}_4$ part of the interaction, which is not captured by the generalized pseudopotentials. As a result, the angular momentum scattering behavior of $V_{\boldsymbol{r}_1- \boldsymbol{r}_2, \boldsymbol{r}_3, \boldsymbol{r}_4}$ is different than we may expect from the generalized pseudopotentials (allowing for changes to the total angular momentum, for example). 

\section{Invariant nullspace dimension with different single-particle normalization factors} 
\label{sec:supp_F}

In this section, we will show that removing the single particle normalization factors $\mathcal{N}_{\boldsymbol{k}}$ in the Hamiltonian does not change the degeneracy of the nullspace. This implies that the ground state degeneracy is topological regardless of how the single-particle orbitals' normalization factors are changed, as long as they remain physically well defined ($|\mathcal{N}_{\boldsymbol{k}}|^2 >0$). Note that such a proof has previously been discussed in Ref.~\cite{Wang2021, Wang2023}. Here we provide a simple proof to make the paper more self-contained.

Mathematically speaking, we want to prove the following theorem:

\noindent \textbf{Theorem (Basis-invariant nullspace dimension)}
Let $A$ be an $n \times n$ diagonalizable matrix over a field $\mathbb{F}$. Suppose we change from an orthonormal basis to a non-orthonormal basis via a linear transformation. Then the dimension of $\ker(A)$ is the same in both bases.

\textit{Proof.} For a diagonalizable matrix $A \in M_{n}(\mathbb{F})$ (where $\mathbb{F}$ is a field such as $\mathbb{R}$ or $\mathbb{C}$), there exists an invertible matrix $P$ and a diagonal matrix $D$ such that $A = P D \, P^{-1}$. We define the nullspace (or kernel) of $A$, denoted $\ker(A)$, as the set of vectors $v$ for which $A\,v = 0$. The degeneracy of the nullspace corresponds to the dimension of $\ker(A)$, equal to the geometric multiplicity of the eigenvalue $0$.

If the matrix representation $A'$ of $A$ in the new basis is given by $A' \;=\; S^{-1}\,A\,S$ so that $A'$ is similar to $A$ and $S$ is unitary, $A$ and $A'$ share identical algebraic and geometric multiplicities. This is the basis transformation we normally perform in the Hilbert spaces. In particular, for the eigenvalue $\lambda = 0$, the geometric multiplicity is exactly the nullspace dimension
\begin{equation}
 \dim\ker(A) = \dim \ker(A') 
\end{equation}
Then we immediately know that a basis change does not affect the nullspace degeneracy.

Generically, $A'$ is not necessarily similar to $A$, especially when we drop the single particle normalization factors as mentioned in the main text. So we can consider the eigenvalue problem for a generic non-orthonormal basis:
\begin{equation}
    A \, v_{i} =\lambda_i \, O \, v_{i},
    \label{eq:generalized_eig}
\end{equation}
where $O_{mn} = e^{\dagger}_{m} e_{n}$ for basis vectors $\set{e_m}$. Then we can choose an invertible matrix $B$ such that $B^{-1} O B$ is diagonal, with diagonal entries $\{s_j\}, s_j \neq 0$. Then define $C_{ij} \equiv B_{ij}/\sqrt{s_j}$, which can be non-unitary, and one can reduce the problem to solving the eigenvalues of the matrix $D = C^{\dagger} A C$. The existence and the number of zero modes remain unaffected by the matrix $S$ because $\forall v_{n}, A\, v_{n} = 0 \Longleftrightarrow D (C^{-1} v_{n}) = 0$. Thus the number of zero modes is insensitive to how the wavefunctions are normalized, ensuring that the system's zero-energy state degeneracy is preserved.

In our case, the normalization factor $\mathcal{N}_{\boldsymbol{k}}$ does not destroy the orthogonality in the many-body basis, so after removing all the $\mathcal{N}_{\boldsymbol{k}_i}$, the zero-energy states in different bases become identical to each other.

\section{Numerical results of modified interactions in \textit{t}MoTe${}_2$ systems} 
\label{sec:supp_G}

In this section, we provide a brief introduction to the continuum model of twisted MoTe$_2$ and more numerical results of the ZDS interaction in bosonic and fermionic systems.

By assuming the spin and valley polarization, the continuum model of \textit{t}TMD can be written as:
\begin{equation}
\mathcal{H}_{t \text{TMD}}=\left(\begin{array}{cc}
-\frac{\hbar^2\left(\mathbf{k}-\boldsymbol{k}_{+}\right)^2}{2 m^*}+\Delta_1(\mathbf{r}) & \Delta_{\mathrm{T}}(\mathbf{r}) \\
\Delta_{\mathrm{T}}^{\dagger}(\mathbf{r}) & -\frac{h^2\left(\mathbf{k}-\boldsymbol{k}_{-}\right)^2}{2 m^{*}}+\Delta_2(\mathbf{r})
\end{array}\right)
\end{equation}
where $\boldsymbol{k}_{\pm}$ are the high symmetry points in the moir\'e Brillouin zone. The interlayer hopping $\Delta_{\mathrm{T}}(\mathbf{r})=w\left(1+e^{-i \mathbf{g}_2 \cdot \mathbf{r}}+e^{-i \mathbf{g}_3 \cdot \mathbf{r}}\right)$, with $\mathbf{g}_i$ as the reciprocal lattice vectors. The intralayer potential is $\Delta_{1,2}(\mathbf{r})=2 V \sum_{j=1,3,5} \cos \left(\mathbf{g}_j \cdot \mathbf{r} \pm \psi\right)$. So the whole model is governed by three parameters $(V,\psi,w)$ once the twist angle is fixed. More details about this model can be found in Ref.~\cite{wu2019mote2,reddy23mote2,chong2024mote2}. In this work, we choose the parameters as $(V, \psi, w)$ = ($20.8$meV, $107.7^\circ$, $-23.8$meV) at the twist angle $3.89^\circ$, and for the interaction we use the relative permittivity $\epsilon= 5$.

\begin{figure}[h]
\includegraphics[width=0.96\linewidth]{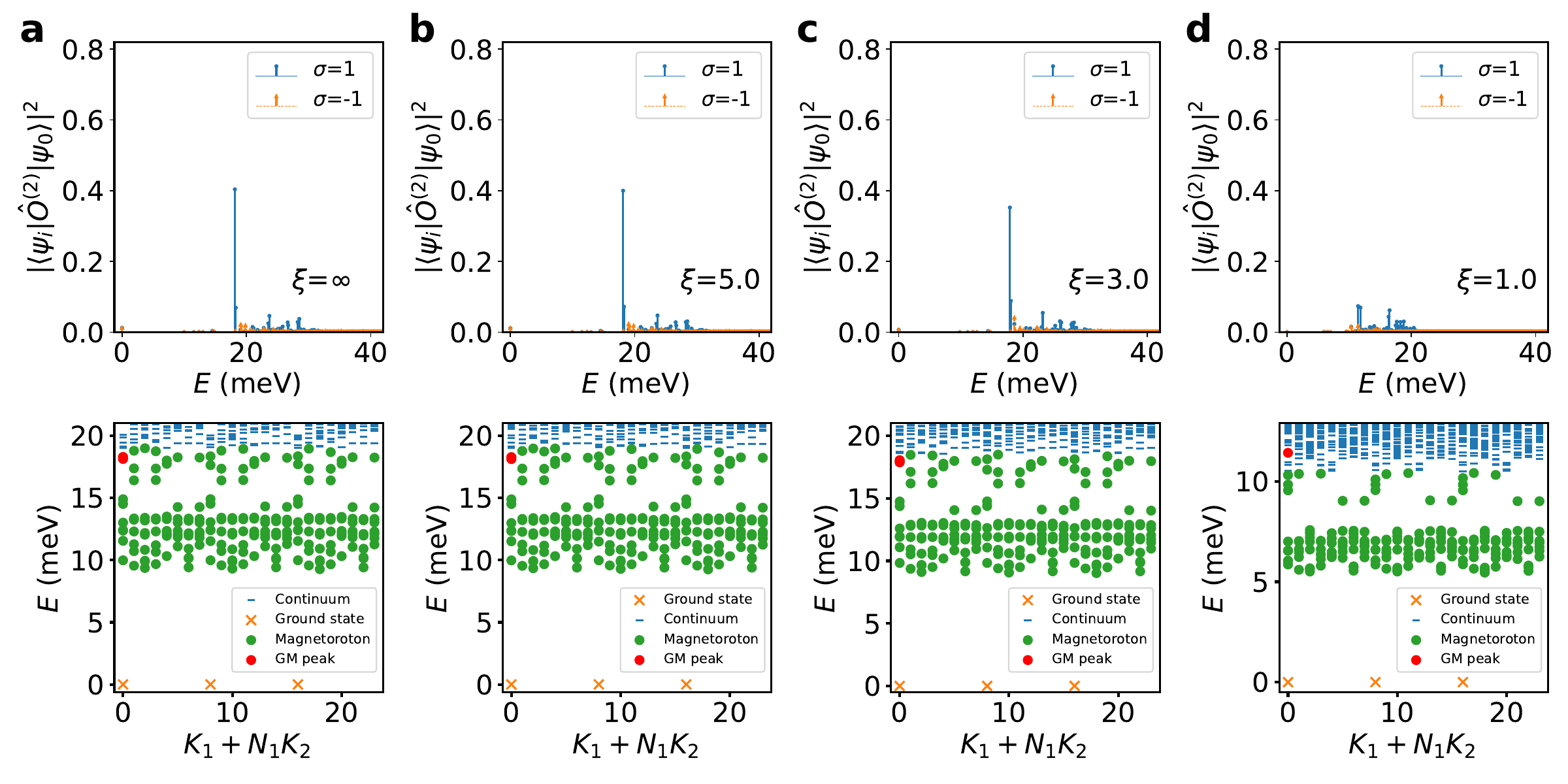}
\caption{\textbf{Spectra and GM spectral functions of screened Coulomb interaction with different screening lengths $\xi$ in fermionic systems}.}
\label{fig_G1}
\end{figure}

In Fig.~\ref{fig_G1}, we show spectral functions of screened Coulomb interactions with different screening lengths $\xi$. When $\xi \rightarrow \infty$, we return to the pure Coulomb interaction. One can see that small screening lengths will bring down the gap and cause a lower GM peak. However, in realistic experiments, the screening effect is insignificant, so additional tuning strategies are required.

\begin{figure}[h]
\includegraphics[width=0.96\linewidth]{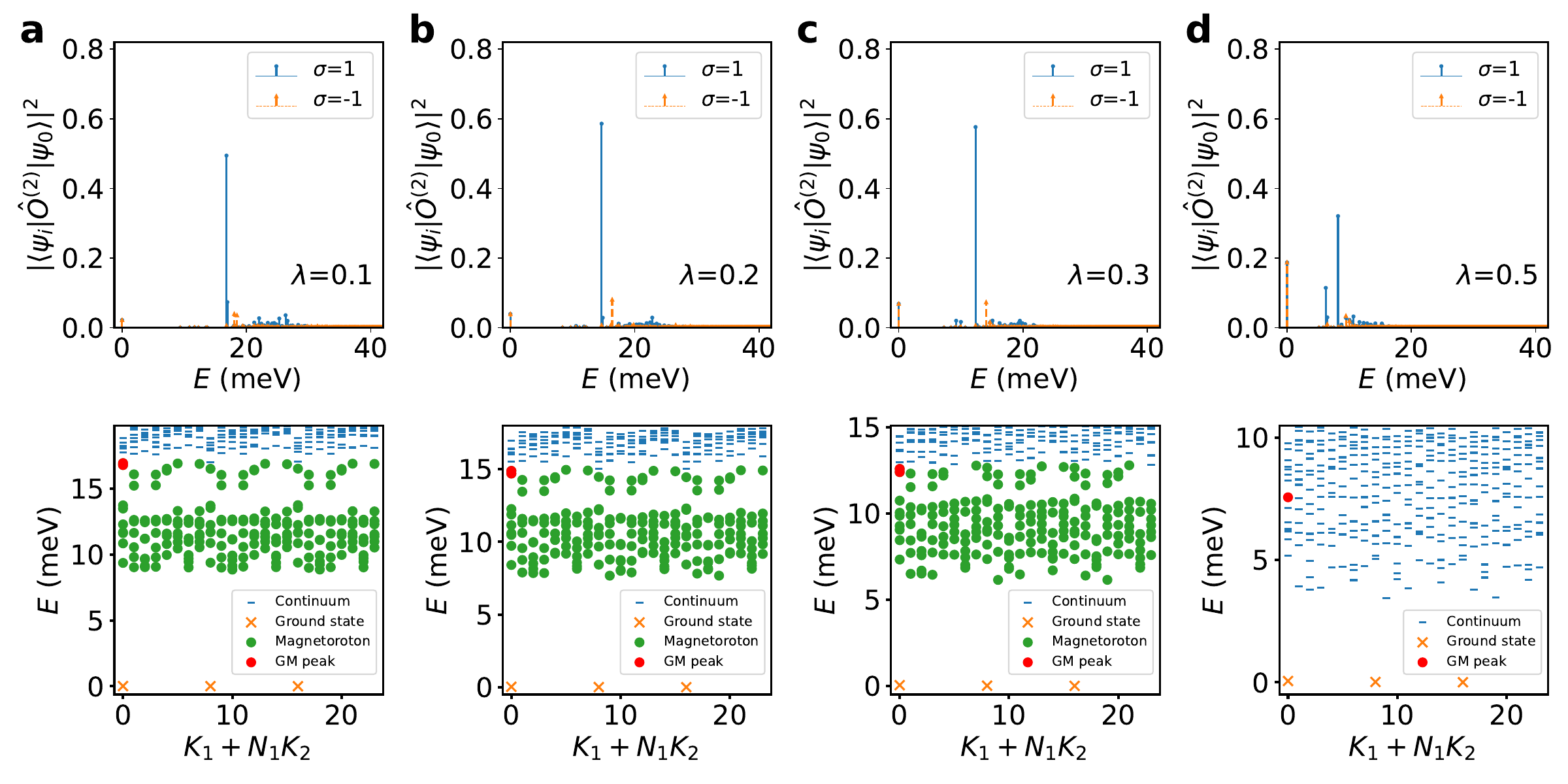}
\caption{\textbf{Spectra and GM spectral functions of $V_{ZDS}$ with different effective thickness $\lambda$ in fermionic systems}. }
\label{fig_G2}
\end{figure}

In Fig.~\ref{fig_G2}, we show the GM spectral functions of the ZDS interactions $V_{ZDS}$ with $N_h=8$ and different effective thickness $\lambda$ in a fermionic \textit{t}MoTe$_{2}$ system at the filling $\nu = -2/3$, where the GM always remains chiral. The GM peak is enhanced with a moderate $\lambda$ but the gaps (both the neutral gap and the gap between the magnetoroton mode and the continuum) keep decreasing with larger $\lambda$. So eventually the GM peak gets lower again, and the FCI phase will get destroyed if $\lambda$ keeps increasing. From the figure, the optimal value of $\lambda$ is observed to be around $0.2$. Note that when we tune the effective thickness to a higher value (such as $\lambda=0.5$ shown in Fig.~\ref{fig_G2}d), the gap between the magnetoroton modes and the continuum will vanish, so the GM is buried deep in the continuum again. If $\lambda$ is even allowed to be arbitrary, one can imagine that the gap will eventually close and the FCI phase will get destroyed. Thus it is important to control the effective thickness within a proper range in experiments. 

\begin{figure}[h]
\includegraphics[width=0.96\linewidth]{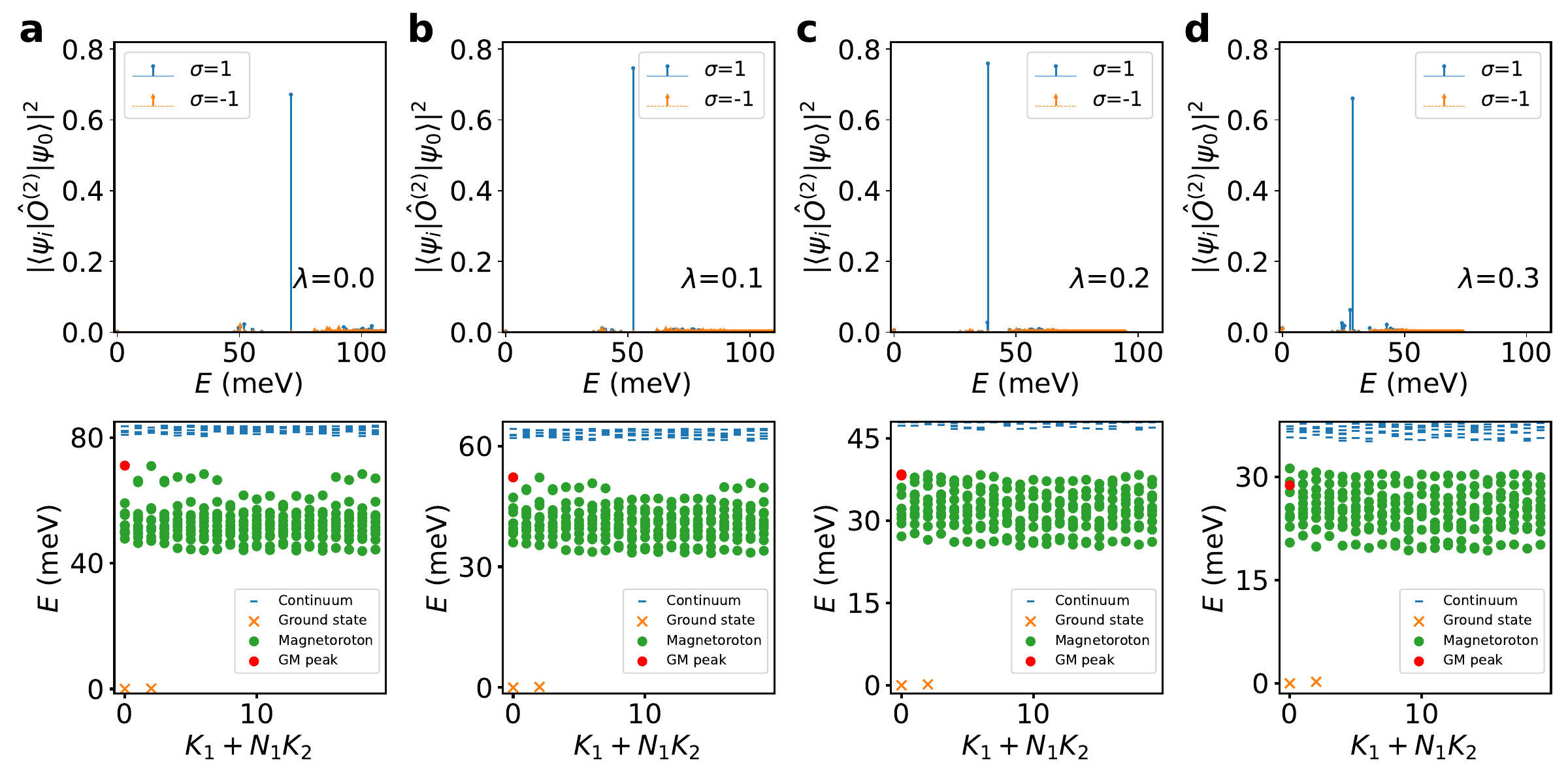}
\caption{\textbf{Spectra and GM spectral functions of $V_{ZDS}$ with different effective thickness $\lambda$ in bosonic systems}. }
\label{fig_G3}
\end{figure}

In Fig.~\ref{fig_G3}, we show spectral functions of the ZDS interactions $V_{ZDS}$ with $N_h = 10$ and different effective thickness $\lambda$ in a bosonic \textit{t}MoTe$_{2}$ system at the filling $\nu = -1/2$. Similarly, the GM peak is chiral and enhanced with a moderate $\lambda$ with the gaps decreasing when $\lambda$ gets large, so one should still control $\lambda$ not to be too large to cause phase transitions. But in this case, even for pure Coulomb, the GM energy is below the continuum (around the middle of the gap between the magnetoroton mode and the continuum). Tuning $\lambda$ also significantly changes the relative position of the GM and the magnetoroton modes, and we can observe that the GM fully enters the magnetoroton modes at around $\lambda = 0.2$. 

\end{document}